\documentclass[prd,twocolumn,tightenlines,showpacs,nofootinbib,superscriptaddress]{revtex4-1}%
\usepackage{bm,dcolumn,amsmath,graphicx,amsfonts,amssymb}

\usepackage{hyperref} 
\usepackage{xcolor}
\definecolor{blue}{rgb}{0.,0.,0.5}   
\hypersetup{ 
	colorlinks,
	linkcolor={blue},
	citecolor={blue},
	urlcolor={blue}
}

\newcommand{\bra}[1]{\ensuremath{\langle #1|}}	
\newcommand{\ket}[1]{\ensuremath{|#1\rangle}}	
\newcommand{\braket}[1]{\ensuremath{\langle #1\rangle}}	
\renewcommand{\v}[1]{\ensuremath{\boldsymbol{#1}}}		
\newcommand{\E}[1]{\ensuremath{\times10^{#1}}}	
\newcommand{\matr}[4]{\ensuremath{\begin{pmatrix}#1&#2\\#3&#4\end{pmatrix}}}	

\newcommand{\psibar}{\ensuremath{\bar\psi}}	
\newcommand{\en}{\ensuremath{\mathcal{E}}}  
\renewcommand{\i}{\ensuremath{\imath}} 
\renewcommand{\j}{\ensuremath{\jmath}} 
\newcommand{\g}{\ensuremath{\gamma}} 
\renewcommand{\a}{\ensuremath{\alpha}}

\renewcommand{\P}{\ensuremath{{P}}}
\newcommand{\C}{\ensuremath{{C}}}
\newcommand{\T}{\ensuremath{{T}}}

\newcommand{\wf}{wavefunction}
 
\newcommand{\smallspace}{\rule{0pt}{2.5ex}}

\begin{document}

\title{Parity-violating interactions of cosmic fields with atoms, molecules, and nuclei:
Concepts and calculations for laboratory searches and extracting limits}  

\author{B. M. Roberts} \email[]{b.roberts@unsw.edu.au}
\author{Y. V. Stadnik} \email[]{y.stadnik@unsw.edu.au}
\author{V. A. Dzuba}   
	\affiliation{School of Physics, University of New South Wales, Sydney 2052, Australia}
\author{V. V. Flambaum} 
	\affiliation{School of Physics, University of New South Wales, Sydney 2052, Australia}
\author{{N. Leefer}}
	\affiliation{Helmholtz Institute Mainz, Johannes Gutenberg University, 55099 Mainz, Germany}
\author{{D. Budker}}	
	\affiliation{Helmholtz Institute Mainz, Johannes Gutenberg University, 55099 Mainz, Germany}
	\affiliation{Department of Physics, University of California at Berkeley, Berkeley, CA 94720-7300, USA}
	\affiliation{Nuclear Science Division, Lawrence Berkeley National Laboratory, Berkeley, CA 94720, USA}

\date{ \today }

\begin{abstract}

We propose methods and present calculations that can be used to search for evidence of cosmic fields by investigating the parity-violating effects, including parity nonconservation amplitudes and electric dipole moments, that they induce in atoms. 
The results are used to constrain important fundamental parameters describing the strength of the interaction of various cosmic fields with electrons, protons, and neutrons.
Candidates for such fields are dark matter (including axions) and dark energy, as well as several more exotic sources described by standard-model extensions. 
Calculations  of the effects induced by pseudoscalar and pseudovector fields  are performed for H, Li, Na, K, Cu, Rb, Ag, Cs, Ba, Ba$^+$, Dy, Yb, Au, Tl, Fr, and Ra$^+$.  
Existing parity nonconservation experiments in Cs, Dy, Yb, and Tl are combined with these calculations to directly place limits on the interaction strength between the temporal component, $b_0$, of a static pseudovector cosmic field and the atomic electrons, 
with the most stringent limit of 
$|b_0^e | < 7\times10^{-15}$ GeV, 
in the laboratory frame of reference, coming from Dy.
From a measurement of the nuclear anapole moment of Cs, and a limit on its value for Tl, we also extract limits on the interaction strength between the temporal component of this cosmic field, as well as a related tensor cosmic-field component $d_{00}$, with protons and neutrons.
The most stringent limits of 
$|b_0^p | < 4\times10^{-8}$ GeV
and
$|d_{00}^p | < 5\times10^{-8}$ 
for protons, and
$|b_0^n | < 2 \times10^{-7}$ GeV
and
$|d_{00}^n | < 2\times10^{-7}$ 
for neutrons (in the laboratory frame) come from the results using Cs. 
Axions may induce oscillating parity- and time-reversal-violating effects in atoms and molecules through the generation of oscillating nuclear magnetic quadrupole and Schiff moments, which arise from $P$- and $T$-odd intranuclear forces and from the electric dipole moments of constituent nucleons. 
Nuclear-spin-independent parity nonconservation effects may be enhanced in diatomic molecules possessing close pairs of opposite-parity levels in the presence of time-dependent interactions.

\end{abstract}
\maketitle

\tableofcontents 

\section{Introduction}

In our recent work~\cite{RobertsCosmic2014}, we proposed methods and presented atomic calculations that can be used for the detection of parity nonconservation (PNC) amplitudes and atomic electric dipole moments (EDMs) that are induced via the interaction of pseudoscalar and pseudovector cosmic fields with atomic electrons and nuclei.
These methods were used to extract limits on the interaction strengths of the temporal component of the pseudovector cosmic field with electrons and protons.
In this work, we describe the method in greater detail, and apply the techniques developed in Ref.~\cite{RobertsCosmic2014} to several more atomic systems and different cosmic fields.
We obtain more accurate limits on the strength of the interaction with protons and extend our previous methods to obtain limits on the strength of the pseudovector cosmic-field interaction with neutrons, by taking into account nuclear many-body effects, which were recently considered in Ref.~\cite{StadnikNMB2014}. We also extend our methods to obtain limits on the interactions of a related tensor cosmic field with protons and neutrons.

One of the most important unanswered questions in fundamental physics today is the so-called strong {\C\P} problem.
This refers to the puzzling observation that quantum chromodynamics (QCD) does not appear to violate the combined charge-parity symmetry (\C\P), despite there being no known theoretical reason for its conservation, see, e.g., Refs.~\cite{Weinberg1976,*Weinberg1978,Peccei1977a,*Peccei1977b,Wilczek1978,*Moody1984}.
One compelling resolution to this problem comes from the Peccei-Quinn (PQ) theory, in which an additional global U($1$) symmetry, known as the PQ symmetry, is introduced into the standard model (SM) QCD Lagrangian and is subsequently broken both spontaneously and explicitly \cite{Peccei1977a,*Peccei1977b} (see also \cite{Kim1979,Zakharov1980,*[][{ [Sov. J. Nucl. Phys. {\bf31}, 529 (1980)].}]Zhitnitsky1980,Srednicki1981}). 
The breaking of the PQ symmetry gives rise to a pseudoscalar pseudo-Nambu-Goldstone boson, born from the QCD vacuum. This particle, known as the axion, causes the QCD {\C\P} symmetry breaking  parameter to become effectively zero, thus in principle alleviating the strong {\C\P} problem. 
For more detail on this topic we direct the reader to the review~\cite{Kim2010} (see also~\cite{PDG2012,Brambilla2014}).

Another crucial outstanding problem in modern physics is the question of dark matter, specifically cold dark matter (CDM).
The astrophysical evidence for the existence of dark matter is overwhelming,
see, e.g., Refs.~\cite{Bertone2005,Agnese2013};
however,  its composition is not known. 
There have been many suggestions put forward that attempt to provide a theoretical framework for dark matter, though no single theory is a clear leading candidate, see, e.g., Refs.~\cite{Bertone2005,PDG2012}. 
What we do know is that the matter-energy content of the universe is dominated by CDM ($\sim23\%$) and dark energy ($\sim73\%$), see, e.g., Ref.~\cite{Spergel2007}.  
Dark energy is proposed to account for the observed accelerating expansion of the Universe~\cite{Riess1998,Perlmutter1999}. 
Even less is known about the composition of dark energy than of CDM.

The axion, since emerging as a compelling solution to the strong {\C\P} problem, has in fact been identified as a promising CDM candidate. 
Axions may constitute a large fraction of the CDM in the observable universe.
Thus axions, if detected, would have a real potential to resolve both the CDM and strong {\C\P} problems, and their detection would provide a great step forward in our understanding of the physical world.
The decay of supersymmetric axions to produce axions may also provide a possible explanation for dark radiation~\cite{Jeong2012,graf2013axions,*graf2013dark,queiroz2014rich}.
Many methods have been proposed and applied to the search for axions; for a review we direct the reader to Refs.~\cite{Rosenberg2000,Kim2010,PDG2012,Kawasaki2013,Baer2014}.

Scalar and pseudoscalar cosmic fields (e.g.~the Higgs and axion fields) have a strong theoretical underpinning.
As well as these, many other background fields are invoked by theories which extend beyond the {SM}, for example, supersymmetric theories and string theory. 
Many of these fields, including vector, pseudovector, and tensor fields, have been conveniently parametrized in the form of the so-called Standard Model Extension (SME)~\cite{Colladay1997,Colladay1998,Kostelecky1999}.
In this work, we focus particularly on the temporal component of the background pseudovector field, which leads to parity-violating effects in atoms.
Limits on the spatial components (which lead to parity-even effects) of this cosmic field have been extracted for the interaction with electrons, protons, and neutrons; see, e.g., Ref.~\cite{Kostelecky2011a,*Kostelecky2014} and references therein.

The prospect that atomic systems could be used as a probe for dark matter, axions, and other cosmic fields has been considered in the literature, see, 
e.g., Refs.~\cite{Pospelov2008,Bolokhov2008,Graham2011,[][{ [Proceedings of the Sixth Meeting on CPT and Lorentz Symmetry, Indiana University, Bloomington, June 2013].}]LeeferCPT2013,
Stadnik2014,Budker2014,Derevianko2010,Karshenboim2011,
Pospelov2013,Derevianko2013,RobertsCosmic2014}.  
While the effects induced in atoms by such a cosmic field may be small, the advantage of using atoms is that {atomic physics} methods are highly advanced, and both the experimental and theoretical accuracy, and hence sensitivity, can be high.

The existence of a cosmic field that interacts with electrons via parity-violating interactions can contribute to the mixing of opposite-parity atomic states, leading to parity-violating effects in atoms.
Parity nonconservation amplitudes are parity-violating $E1$ transitions between two states of the same nominal parity.
They are generated by parity-violating forces; 
in the conventional case, these include $Z^0$-boson exchange between the electrons and nucleons and the electromagnetic interaction of the electrons with parity-violating nuclear moments that are borne by parity-violating forces inside the nucleus.
Complementary to direct tests performed at high energy (e.g.~at CERN), measurements of PNC amplitudes and EDMs in atoms are relatively inexpensive low-energy tests of the standard model, see, e.g., Refs.~\cite{Khriplovich1991,GingesRev2004,DzubaReview2012}.
The PNC amplitude of the $6s$-$7s$ transition in cesium is the most precise atomic test of the electroweak theory to date.  
This precision is due to the highly accurate (0.35\%) measurements~\cite{Wieman1997} (see also~\cite{Bouchiat1982,Watts1985,Lintz2007}), and the almost equally accurate atomic calculations (0.5\%) that are needed for their interpretation~\cite{DzubaCPM1989plaPNC,DzubaCPM1989plaEn,*DzubaCPM1989plaE1,
Blundell1992,KozlovCs2001,DzubaCs2002,Porsev2009,*Porsev2010,OurCsPNC2012}.
These studies show that the observed value of the nuclear weak charge for cesium-133 agrees with the SM prediction to the 1.5$\sigma$ level~\cite{Wieman1997,OurCsPNC2012,Blunden2012}.

In addition to inducing PNC effects and EDMs, cosmic fields that interact with standard-model fermions can give rise  to other fascinating phenomena.
In the case of axions, this includes the axio-electric effect~\cite{Avignone1987,Pospelov2008,Avignone2009a,Avignone2009b,Derevianko2010,
DzubaPRD2010,[][{ [Pis'ma Zh. Eksp. Teor. Fiz. {\bf95}, 379 (2012)].}]Derbin2012,Derbin2013}, nuclear anapole moments, and spin-gravity and spin-axion momentum couplings in atomic, molecular, solid-state and nuclear systems \cite{Kimball2013,Venema1992,Stadnik2014,Graham2013}.
All of these effects can in principle be observed. 
A general pseudoscalar cosmic field need not necessarily be restricted to an axionic one; dark energy and other exotic fields are also possibilities. 
We therefore present the atomic-structure calculations separately from any field parameters, to avoid any model dependence.

In Sec.~\ref{sec:theory}, we first show that a static pseudoscalar cosmic field cannot give rise to observable {\P}-odd effects in atoms in the lowest order, and then present the necessary theory and derive expressions for the PNC effects and EDMs induced in atoms and nuclei by pseudoscalar and pseudovector cosmic fields.
We also note that axions may induce oscillating {\P}- and {\T}-odd effects in molecules through the generation of oscillating nuclear magnetic quadrupole moments, which arise from  {\P}- and {\T}-odd intranuclear forces and from the EDMs of constituent nucleons.
Nuclear-spin-independent PNC effects may be enhanced in diatomic molecules possessing close pairs of opposite-parity levels in the presence of time-dependent interactions, in contrast to the static case, where only nuclear-spin-dependent PNC effects are enhanced.
We go on in Sec.~\ref{sec:calcs} to present the methods used for our {\sl ab initio} relativistic atomic calculations for pseudovector and dynamic pseudoscalar cosmic-field--induced PNC amplitudes and atomic EDMs for
a number of neutral atoms and ions.
These calculations are necessary for determining, or placing limits upon, important pseudoscalar and pseudovector cosmic-field parameters, in conjunction with appropriate experimental data. 
In Sec.~\ref{sec:results}, we present the results of our atomic calculations and combine these with existing PNC experiments in cesium, thallium, ytterbium, and dysprosium to give limits on the interaction strengths of static pseudovector and tensor cosmic fields with electrons, protons, and neutrons. 
We also discuss possible systems for experimentally obtaining limits on the interaction strengths of dynamic cosmic fields with standard model fermions.

\section{Theory}
\label{sec:theory}

\subsection{Parity-violating interactions of fermions with cosmic fields} 

Except where explicitly noted, we use natural units, $\hbar=c=1$, throughout.
In this work, we consider two distinct sources of cosmic fields.
Pseudoscalar (PS) fields, such as axions, are described by the Lagrangian density
\begin{equation}
\mathcal{L}^{\rm PS} =
 -i \zeta m_f\, \phi  \,  \psibar \gamma^5  \psi 
+ \eta (\partial_\mu\phi)\, \psibar \gamma^\mu\gamma^5\psi
\label{eq:lps},
\end{equation} %
where $\zeta$ and $\eta$ are dimensionless constants quantifying the interaction strength of fermions with the PS cosmic field via a direct and derivative-type coupling, respectively,
$m_f$ is the mass of the fermion in question, $\psi$ is the fermion {\wf} with the Dirac adjoint $\bar\psi \equiv \psi^\dagger\gamma^0$, and $\g^{\mu}$ (with ${\mu=0,1,2,3}$) and $\g^5=i\g^0\g^1\g^2\g^3$ are the Dirac matrices.

Here, $\phi=\phi(\v{r},t)$ is the dynamic PS field in question. 
In the next section we will see that an interaction of the form (\ref{eq:lps}) with a static field will not lead to any parity-violating effects in atoms in the lowest order.
The field $\phi$ (for example, an axion field or a light pseudoscalar dark-matter field) obeys the Klein-Gordon equation,
$[\partial_\mu\partial^\mu + m^2]\phi=0$.   
We take this field to be classical and real, so that
 \begin{equation}
\phi(\v{r},t) = \cos(\omega_\phi t-\v{p}_\phi\cdot\v{r}+\xi), 
\label{eq:phi}
\end{equation}
where $\v{p}_\phi$ and $\omega_\phi$ are the momentum and energy of the pseudoscalar field particle (e.g.~the axion), respectively, and $\xi$ is a phase factor. 
We have absorbed the amplitude of the field into the constants $\eta$ and $\zeta$. 
With a redefinition of the phase factor at a fixed point in space, we can express this field more simply as 
$\phi(\v{r},t) = \cos(\omega_\phi t)$. 
This is valid so long as the time scale of an experiment is sufficiently short that  the evolution of the $\v{p}_\phi\cdot\v{r}$ term in (\ref{eq:phi}), 
which corresponds to the motion of the observer with respect to the coordinates,  is small compared with the evolution of the $\omega_\phi t$ term over the course of the experiment.  
This will usually be the case, since the typical speed of a PS cosmic field relative to Earth is expected to be $v\sim10^{-3}$, see, e.g., Ref.~\cite{Bertone2005}; a brief discussion of the coherence time is given towards the end of the paper.

We also consider terms from the local Lorentz-invariance-violating SME~\cite{Colladay1997,Colladay1998,Kostelecky1999}:
\begin{equation}
\mathcal{L}^{\rm SME}
= \frac{1}{2}i\psibar\Gamma_\nu\tensor{\partial}^\nu\psi 
- \bar\psi M\psi,
\label{eq:sme}
\end{equation}
where
\begin{align}
M 
&= a_\mu\g^\mu+b_\mu\g^\mu\g^5+\frac{1}{2}H_{\mu\nu}\sigma^{\lambda\mu},
\label{eq:smeM}
\\
\Gamma_\nu 
&= c_{\mu\nu}\g^\mu+d_{\mu\nu}\g^\mu\g^5+e_\nu+if_\nu\g^5+\frac{1}{2}g_{\lambda\mu\nu}\sigma^{\lambda\mu},
\label{eq:smeG}
\end{align}
$\sigma^{\lambda\mu}=i[\g^\lambda,\g^\mu]/2$,
where $[A,B]=AB-BA$ is the commutator,
and $A\tensor{\partial}^\nu B =A({\partial}^\nu B) -   ({\partial}^\nu A)B$, where the derivatives act on the {\wf}s only (not the fields).
The relativistic interaction Hamiltonians due to Eqs.~(\ref{eq:smeM}) and (\ref{eq:smeG}) are
\begin{align}
\hat h^M &= a_0 + a_j\g^0\g^j + b_0\g^5 + b_j\g^0\g^j\g^5 
\notag\\&\phantom{=~}
+ iH_{0j}\g^j + \frac{1}{2}H_{jk}\epsilon^{jkl}\g_l\g^5
\label{eq:smehM}
\end{align}
and
\begin{align}
\hat h^\Gamma &=
c_{00}\g^0\g^jp_j 
-(c_{0j}+c_{j0})p^j
-c_{jk}\g^0\g^jp^k
-mc_{00}\g^0
\notag\\&\phantom{=}
+d_{00}\g^0\g^j\g^5p_j
-(d_{0j}+d_{j0})\g^5p^j
-d_{jk}\g^0\g^j\g^5p^k
\notag\\&\phantom{=}
-m_fd_{j0}\g^j\g^5
-m_fe_0
-e_j\g^0p^j
-if_j\g^0\g^5
\notag\\&\phantom{=}
-\epsilon^{jkl}g_{j00}\g_l\g^5p_k
+i(g_{j0k}+g_{jk0})\g^jp^k
\notag\\&\phantom{=}
+\frac{1}{2}\epsilon^{jkl}g_{jkm}\g_l\g^5p^m
-\frac{1}{2}m_f\epsilon^{jkl}g_{kl0}\g^0\g_j\g^5,
\label{eq:smehG}
\end{align}
respectively~\cite{Kostelecky1999a}. (Also see Ref.~\cite{Kostelecky1999a} for a derivation of the nonrelativistic form of the above Hamiltonian.)
In the above equations, the Lorentz indices are separated into their time and space components, with Latin characters $j,k,l,m$ running 1 through 3,  and $\g^a=-\g_a$. We use the standard (+$-$$-$$-$) metric, and a summation over repeated indices is assumed.

We note that interactions of cosmic fields with fermions are not limited to those described by the SME Lagrangian~(\ref{eq:sme}).
For example, dimension-five operators that are linear in the electromagnetic gauge-field strength, see, e.g.,~\cite{Bolokhov2005,Bolokhov2008}, can produce static electric dipole moments of fundamental particles~\cite{Bolokhov2008}, and contribute to the splitting of the magnetic dipole moments of fermions and their antifermion partners~\cite{Bolokhov2005,StadnikAMM2014}.

\subsection{Interaction of electrons with pseudoscalar and pseudovector cosmic fields}

\begin{figure}
	\begin{center}
		\includegraphics[width=4cm]{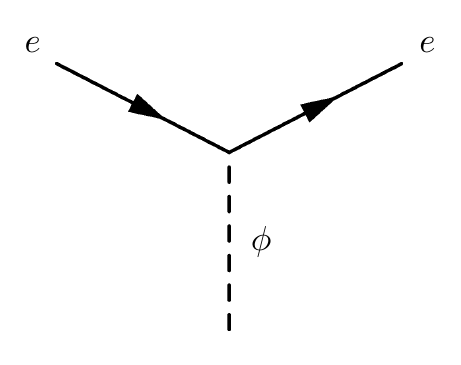}
		\caption{Fundamental vertex for the interaction of an electron with a pseudoscalar cosmic field $\phi$ via the coupling (\ref{eq:lps}).}
		\label{fig:scalar_FD}
	\end{center}
\end{figure}

The direct PS interaction [first term of the right-hand side of (\ref{eq:lps})], and the  time-derivative part of the derivative-type PS interaction [second term on the right-hand side of (\ref{eq:lps})], lead to interaction Hamiltonians of the form 
\begin{equation}
\hat h_{i\g^0\g^5}^{\rm PS}
= i \zeta m_f \cos(\omega_\phi t) \g^0\g^5,
\label{eq:H-ig0g5}
\end{equation}
and
\begin{equation}
\hat h_{\g^5}^{\rm PS} 
= \eta \omega_\phi  \sin(\omega_\phi t) \g^5,
\label{eq:H-g5}
\end{equation}
which we shall refer to as the PS $i\g^0\g^5$ and the PS  $\g^5$ interactions, respectively\footnote{Note that the `$\g^5$' interaction appears as $\g^0\g^5$ in the Lagrangian (and visa-versa); this possibly confusing notation stems from the extra $\g^0$ in $\bar\psi=\psi^\dagger\g^0$.}.
The fundamental vertices for the interactions (\ref{eq:H-ig0g5}) and (\ref{eq:H-g5}) are represented by the same Feynman diagram (presented in Fig.~\ref{fig:scalar_FD}).
Interactions of this form with atomic electrons will  manifest themselves as oscillating contributions to PNC amplitudes and atomic EDMs.

\begin{figure}
	\begin{center}
		\includegraphics[width=4cm]{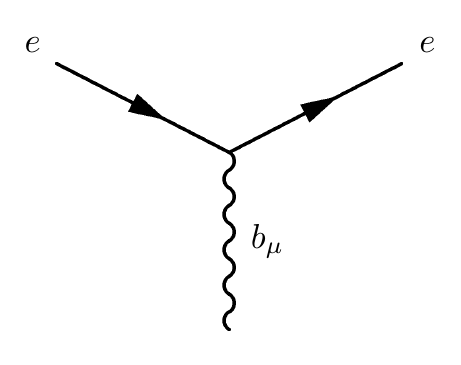}
		\caption{Fundamental vertex for the interaction of an electron with a pseudovector cosmic field $b_\mu$ via the coupling (\ref{eq:L-static}).}
		\label{fig:PV_FD}%
	\end{center}
\end{figure}

It is also possible for parity-violating interactions of electrons with a cosmic field to produce static PNC effects in atoms.
For this, we consider the Lagrangian corresponding to the interaction of electrons with the pseudovector (PV) field, $b_\mu$ (Fig.~\ref{fig:PV_FD}):
\begin{align}
\mathcal{L}_{\rm \gamma^5}^{\rm PV} 
&=  b_\mu\bar\psi \gamma^\mu\gamma^5\psi \notag \\
&=   b_0\psi^\dagger \gamma^5\psi  
	-\v{b}\cdot \psi^\dagger \v{\alpha}\gamma^5\psi,
\label{eq:L-static}
\end{align}
where $\v{\alpha}=\g^0\v{\g}$, and we have absorbed the strength of the interaction into the definition of the field $b_{\mu}=(b_0,-\v{b})$.
The temporal-component term of this coupling leads to the interaction Hamiltonian
\begin{equation}
\hat h_{\g^5}^{\rm PV}
=  \, b_0(t) \g^5,
\label{eq:H-static}
\end{equation}
which could be either static [$b_0(t)=b_0$] or dynamic [$b_0(t)=b_0\sin(\omega_b t)$] (the choice of phase here is entirely arbitrary, and is chosen for later convenience). 
We refer to this interaction as either the static or dynamic PV $\g^5$ interaction.
In the dynamic case, the effects of (\ref{eq:H-static}) will mimic those of (\ref{eq:H-g5}).
In the static case, however, they will mimic the conventional nuclear-spin-independent (NSI) PNC signal induced by $Z^0$-boson exchange between the nucleus and electrons, described by the Hamiltonian
\begin{equation}
\hat h_{Q_W} = \frac{G_F}{2\sqrt{2}}Q_W \rho(\v{r})  \gamma^5,
\label{eq:hqw}
\end{equation}
where $G_F=1.166\times10^{-5}$ GeV$^{-2}$
is the Fermi weak constant, $Q_W$ is the nuclear weak charge and $\rho$ is the normalized nucleon density.
In the standard model, $Q_W$ is approximately equal to the number of neutrons in the nucleus.

The spatial-derivative component terms in (\ref{eq:lps}), and the spatial component terms in (\ref{eq:L-static}), lead to interaction terms of the form 
$\v{\sigma} \cdot \v{B}_{\textrm{eff}}$, 
where $\v{\sigma}$ is the spin of a SM fermion and $\v{B}_{\textrm{eff}}$ is an effective magnetic field due to the momentum of the PS or PV cosmic field, and thus give no parity-violating effects. 
The best current limits on such static interactions of a cosmic field with electrons, protons, and neutrons, using the notation of the SME parametrization~\cite{Kostelecky1999}, are: 
$|\tilde{b}_{X}^e|< 1.3 \times 10^{-31}$ GeV, 
$|\tilde{b}_{Y}^e|< 1.3 \times 10^{-31} $ GeV~\cite{Heckel2006,*Heckel2008}, 
$|\tilde{b}_{\perp}^p|< 1.6 \times 10^{-33} $ GeV~\cite{StadnikNMB2014} 
and 
$|\tilde{b}_{\perp}^n| < 8.4 \times 10^{-34} $ GeV~\cite{Allmendinger2014} (see also \cite{StadnikNMB2014}), 
respectively, where the subscripts denote the field components in the Sun Centered Celestial Equatorial Frame (SCCEF). 
Here and throughout this work, the superscripts $e$, $p$, and $n$ denote the particle species: electron, proton, and neutron, respectively.
For further details on the broad range of experiments performed in this field and a brief history of recent developments in the improvement of these limits, we refer the reader to
Refs.~\cite{Berglund1995,Kostelecky1999,Hou2003,Cane2004,Heckel2006,*Heckel2008,Altarev2009,
Brown2010,Gemmel2010,Peck2012,Allmendinger2014}.
A comprehensive list of the limits extracted for the various interaction constants has been compiled in Ref.~\cite{Kostelecky2011a,*Kostelecky2014}. 
Indirect limits have been obtained for the SME parameter $\tilde{b}_{T}^e$  through linear combinations of several SME parameters, constrained at the level of 
$\sim 2 \times 10^{-27}$ GeV~\cite{Heckel2006,*Heckel2008}. 
Indirect limits have also been obtained for the SME parameter $\tilde{b}_{T}^n$ \cite{Cane2004}. 
In the present work, we consider the extraction of direct limits on the 
{\P}-odd effects induced by the 
temporal component of the field, $b_0$ [as defined in Eq.~(\ref{eq:H-static})], for electrons, protons and neutrons, which are complementary to the limits derived from {\P}-even fermion effects discussed above. 
We will not be considering the cosmic-field--induced interaction $\v{\sigma} \cdot \v{B}_{\textrm{eff}}$ further in this work, but note that such an interaction can also be sought in an oscillatory form (see, e.g., Refs.~\cite{Stadnik2014,Graham2013}).

Note that any effective Hamiltonian that is proportional to the $\gamma^5$ or $i\gamma^0\gamma^5$ matrices will lead to a mixing of opposite-parity states in atoms and thus could contribute to parity nonconserving amplitudes. 
In this sense, the calculations provided in this work are general, and can be applied to any source leading to an interaction in the above forms.

The matrix elements of the $\gamma^5$ and $i\g^0\g^5$ operators are not entirely independent of one another.
Considering the  relativistic Hamiltonian for an $N$ electron atom of nuclear charge $Z$ in the presence of electrostatic interactions,
\begin{equation}
 \hat H  = \sum_{\i=1}^N \left[ \v{\alpha}_\i\cdot\v{p}_\i+ m_e (\g^0_\i -1) - \frac{Ze^2}{r_\i}+\sum_{\j<\i} \frac{e^2}{r_{\i\j}}\right],
\label{eq:H-electrostat}
\end{equation}
where $\v{p}_\i$ is the relativistic momentum of the $\i$th electron, $r_{\i\j}=|\v{r}_\i-\v{r}_\j|$, and $e=|e|$ is the elementary charge,
the two operators in question are related via the useful identity 
\begin{equation}
i\gamma^0_k\gamma^5_k = \frac{i}{2m_e}[\hat H,\gamma^5_k]     
\label{eq:comm}
\end{equation}
 (proved below), from  which it follows that   
\begin{equation}
\bra{b}i\gamma^0_k\gamma^5_k\ket{a} = \frac{i}{2m_e}(E_b-E_a)\bra{b}\gamma^5_k\ket{a},
\label{eq:relation}
\end{equation}
where the states $a$ and $b$ are eigenstates of the atomic Hamiltonian (\ref{eq:H-electrostat}) with eigenvalues $E_a$ and $E_b$ respectively.
Note that for the standard choice of angular {\wf}s, the matrix elements of the $i\gamma^0\gamma^5$ operator are real and hence  symmetric, whereas the $\gamma^5$ operator gives rise to imaginary matrix elements, and are antisymmetric. Equation~(\ref{eq:relation}) maintains this symmetry.
To prove the relation in the case of the electrostatic Hamiltonian (\ref{eq:H-electrostat}), note that the commutator in Eq.~(\ref{eq:comm})  reduces to
\begin{align}
[\hat H,\gamma^5_k] 
&= \sum_{\i}\left( 
 	[ \v{\alpha}_\i,\gamma^5_k]\cdot\v{p}_\i
	+m_e[\gamma^0_\i,\gamma^5_k]
	\right)\notag\\
&=2m_e\gamma^0_k\gamma^5_k.
\end{align} 
We have made use of the relation $\{\gamma^\mu,\gamma^5\}=0$ for $\mu$=0,1,2,3 ($\{x,y\}=xy+yx$ is the anti-commutator). 
This relation holds equally well if we had used the Hartree-Fock Hamiltonian (including core polarization) in place of the `exact' Hamiltonian (\ref{eq:H-electrostat}). In that case, the many-body {\wf}s and energies that appear in Eq.~(\ref{eq:relation}) would be replaced by their single-particle counterparts.

The atomic PNC amplitude can then be written as
\begin{equation}
E_{\rm PNC}^{a\to b} = \sum_k\bra{\widetilde{b(t)}}\v{d}_k \ket{\widetilde{a(t)}},
\label{eq:epnc}
\end{equation}
where $\v{d}_k =-e\v{r}_k$ is the operator of the electric dipole ($E1$) interaction, and
$\ket{\widetilde a}=\ket{a} + \ket{\delta a}$ 
is the perturbed {\wf} associated with the atomic state $a$, with $\ket{a}$ the unperturbed {\wf}, and $\ket{\delta a}$ is the correction to the {\wf} due to the PNC interactions (\ref{eq:H-ig0g5}), (\ref{eq:H-g5}) or (\ref{eq:H-static}).
Likewise, the induced atomic EDM can be expressed as
\begin{equation}
d_{\rm EDM}^{a} = \sum_k\bra{\widetilde{a(t)}}\v{d}_k \ket{\widetilde{a(t)}}.
\label{eq:dedm}
\end{equation}

\subsection{Interaction of atomic electrons with a static pseudoscalar field and other SME terms}

Before we present the formulas for $\ket{\widetilde a}$, we discuss briefly the effects of a possible static pseudoscalar interaction, and show that such an interaction cannot give rise to observable {\P}-odd amplitudes in atoms in the lowest order (though note that a static pseudo{vector} field can).
To see this for a derivative-type coupling, note that the time derivative in the interaction Lagrangian density (\ref{eq:lps}) vanishes for a static field $\phi$.
The spatial derivative terms in (\ref{eq:lps}) lead only to {\P}-even effects, since they cannot lead to mixing of opposite parity atomic states.

 To see this for the direct pseudoscalar coupling [first term on the right-hand side of (\ref{eq:lps})], we 
prove a general relation that states that any static interaction Hamiltonian, $\hat h$, that can be expressed in the form
\begin{equation}
\hat h = [\hat H,\hat o],
\label{eq:comH}
\end{equation}
where $\hat H$ is the atomic Hamiltonian (\ref{eq:H-electrostat}), will not give rise to any electromagnetic amplitudes, which have the form 
$j_\mu A^\mu = \psi^\dagger_b(A^0 + \v{\alpha}\cdot\v{A})\psi_a$, 
 in atoms, where $A^\mu=(A^0,\v{A})$ is the photon field.
This will hold so long as the commutator $[ A^0+\v{\alpha}\cdot\v{A},\hat o]=0$.

Using time-independent perturbation theory, the {\wf},
$\ket{\tilde{a}} =\ket{a}+\ket{\delta a}$, 
perturbed to first-order by the interaction $\hat h$ can be written as 
\begin{align}
 \ket{\tilde{a}}  
&=    \ket{a} +\sum_{n}  \frac{\ket{n} \bra{n} \hat h \ket{a}}{E_{a}-E_{n}}\notag\\
&= \ket{a} - \hat o\ket{a},
\label{eq:dpsi-static}
\end{align}
and
\begin{align}
 \bra{\tilde{a}}  &=  \bra{a} + \bra{a}\hat o,
\end{align}
where, with the use of the relation (\ref{eq:comH}), the energy denominators cancel, and the summation is reduced to unity by closure. 
One can also check that $\ket{\tilde a}$ in (\ref{eq:dpsi-static}) is the solution of the Dirac equation with the perturbation (\ref{eq:comH}).
Hence, the correction induced by the static interaction $\hat h$ to any general electromagnetic interaction is reduced to
\begin{equation}
\bra{{b}}(A^0+\v{\alpha}\cdot\v{A})\ket{\delta{a}}+
\bra{\delta{b}}(A^0+\v{\alpha}\cdot\v{A})\ket{{a}}
=\bra{b}[ A^0+\v{\alpha}\cdot\v{A},\hat o]\ket{a}
\label{eq:relation2}.
\end{equation}
There are thus no corrections to electromagnetic amplitudes if the commutator in (\ref{eq:relation2}) is equal to zero.
Note also that any operator satisfying Eq.~(\ref{eq:comH}) automatically has no diagonal matrix elements and has null expectation values for an energy eigenstate.

In the case of PNC amplitudes and atomic EDMs, including (\ref{eq:epnc}) and (\ref{eq:dedm}), the relevant electromagnetic interaction operator is the $E1$ operator, $\v{d}$.
For the static pseudoscalar interaction [Eq.~(\ref{eq:H-ig0g5}) with $\omega_\phi=0$], $\hat h=i\gamma^0\gamma^5$, and from Eq.~(\ref{eq:comm}), $\hat o\propto\gamma^5$. 
Since $[\g^5,\v{r}]=0$, the static pseudoscalar field does not give rise to any observable {\P}-odd transitions or EDMs in atoms in the lowest order.
Also, since the commutator is equal to zero,  the correction to the {\wf} (\ref{eq:dpsi-static}) does not contribute to the  Dirac charge or current densities $j_{\mu}$.

The PV field (\ref{eq:L-static}) and the dynamic PS fields (\ref{eq:lps}) will be examined in detail in the rest of the paper.
Here we turn our attention briefly to some of the other fields in the SME and discuss what possible parity-violating effects they could give rise to in atomic systems.

The $a_\mu$ term in the SME Lagrangian (\ref{eq:smeM}) is equivalent to interaction with a constant vector potential and does not give rise to observable effects in atoms. 
It is also easy to check directly that 
$a_\mu \v{\alpha}=i[\hat H, a_\mu \v{r}]$
 and that therefore constant $a_\mu$ contributions vanish in atoms.
We note, however, that due to the {\C\P\T}-odd charge nonconservation --- the fact that these fields may couple to different fermion species with different interaction strengths or charges ---  interactions involving more than one fermion species, such as in particle decays, may be affected.

The $e_j$ interaction term can be expressed as 
$\hat h_{e_j} = \g^0\v{e}\cdot\v{p}$, 
which gives no effects in atoms. This can be demonstrated as follows.
Using the relations 
$[\hat H,\v{\g}]=-2\g^0\v{p} + 2m_e\v{\alpha}$,
and
$[\hat H,\v{r}]=-i\v{\alpha}$,
which hold for the atomic Hamiltonian (\ref{eq:H-electrostat}), this term can be expressed as
\begin{equation}
\hat h_{e_j} = \v{e}\cdot [\hat H, {im_e}\v{r} - \frac{1}{2}\v{\g}],
\end{equation}
which is in the form of Eq.~(\ref{eq:comH}) and hence gives no atomic effects due to Eqs.~(\ref{eq:dpsi-static}) and (\ref{eq:relation2}).

The $d_{00}$ and $d_{jk}$ terms in the SME (\ref{eq:smehG}) lead to interaction Hamiltonians proportional to 
$d_{00}\v{\hat \Sigma}\cdot \v{p}$, 
and
$d_{jk}{\hat \Sigma}^j {p}^k$, respectively,
where 
$\v{\hat\Sigma} = \matr{\v{\sigma}}{0}{0}{\v{\sigma}}$
 is the Dirac spin matrix. 
These terms both lead to parity-violating effects in atoms.
We consider the $d_{00}$ term for interactions with nucleons in Sec.~\ref{sec:nucleons}.
In the nonrelativistic limit, this term will not lead to any atomic effects via an interaction with electrons, since in this limit it can be expressed 
$im_e[\hat H_{\rm NR},\v{\sigma}\cdot\v{r}]$, where $\hat H_{\rm NR}$ is the nonrelativistic Schr\"odinger Hamiltonian.
The $H_{0j}$, $g_{jkm}$, and $g_{j00}$ terms in (\ref{eq:smehM}) and (\ref{eq:smehG}) also lead to parity-violating effects in atoms, though we do not consider these in this work.

Many of the terms in the SME Lagrangian (\ref{eq:sme}) are proportional to $\v{p}$ in the nonrelativistic limit and, because of the relation $\v{p}=im_e[\hat H_{NR},\v{r}]$, give no atomic effects in this limit.
The $c_{0j}$ terms, which in the nonrelativistic limit scale as $\v{p}$, also produce {\P}-odd effects due to relativistic corrections.
Also, they introduce direction and frame dependent anisotropies in the electron energy-momentum relation~\cite{Cane2004,Hohensee2013,LeeferCPT2013}.

The other terms in Eqs.~(\ref{eq:smehM}) and (\ref{eq:smehG}) give rise to {\P}-even interactions, and do not contribute to atomic parity-violating effects.
These terms do, however, contribute to other interesting phenomena, such as bound-state energy shifts and modulations in clock transition frequencies. 
For more information and detailed discussions of many of these terms, see, e.g., 
Refs.~\cite{Colladay1997,Colladay1998,Kostelecky1999,
Kostelecky2011a,*Kostelecky2014}.

\subsection{Perturbed {\wf}s and formulas for the atomic PNC amplitudes and EDMs}

To analyze the dynamic effects, we apply first-order time-dependent perturbation-theory (TDPT) with a slow turn-on of the perturbation (see, e.g., Ref,~\cite{Stadnik2014} for further details), and find that the perturbed {\wf} corresponding to the unperturbed atomic state $\ket{a}$ due to the considered dynamic interactions is given by
\begin{equation}
\ket{\widetilde{a(t)}} = \ket{a} + \sum_n c_n^{(a)}(t)\ket{n},
\label{eq:dpsi}
\end{equation}
where
\begin{align}
c_n^{(a)}(t) 
=&  \frac{\sum_\i\bra{n}\hat V_\i\ket{a}}{(E_a-E_n)^2-\omega_\phi^2}
 \left[-i\partial_t f(t) + (E_a-E_n)f(t)\right].
\label{eq:cnt}
\end{align}
Here, 
{$f(t)=\eta\omega_\phi\sin(\omega_\phi t)$} and $\hat V=\g^5$ when we consider the PS  $\g^5$ interaction (\ref{eq:H-g5}),
{$f(t)=\zeta m_e \cos(\omega_\phi t)$} and {$\hat V=i\g^0\g^5$} when we consider the PS $i\g^0\g^5$ interaction (\ref{eq:H-ig0g5}), 
and
{$f(t)=b_0\sin(\omega_b t)$} and $\hat V=\g^5$ when we consider the dynamic case of the PV  $\g^5$ interaction (\ref{eq:H-static}).
The index $\i$ denotes summation over atomic electrons.
In deriving Eq.~(\ref{eq:cnt}), we have neglected the natural widths of the considered states.
While we do not consider these widths in this work, 
they may affect the phase in (\ref{eq:cnt}) when considering resonance phenomena.

Therefore, the general PNC amplitude can be expressed to first order in the PNC interaction as:
\begin{widetext}
\begin{equation}
E^{a\to b} =  \sum_{n,\i,\j} \left\{ 
\frac{\bra{b}\v{d}_\j\ket{n}\bra{n}\hat V_\i\ket{a}}{(E_a-E_n)^2-\omega_\phi^2} \left[- i\partial_t f(t) + (E_a-E_n)f(t) \right]
+
\frac{\bra{b}\hat V_\i\ket{n}\bra{n}\v{d}_\j \ket{a}}{(E_b-E_n)^2-\omega_\phi^2} \left[ i\partial_t f(t) + (E_b-E_n)f(t) \right]
\right\}.
\label{eq:pnc-general}
\end{equation}
\end{widetext}
Note that Eq.~(\ref{eq:pnc-general}) also applies for induced atomic EDMs, for which the initial and final atomic states are identical.

It is now convenient to make one further approximation, namely that the energy of the field particle is much smaller than the energy separation between all opposite-parity states of interest, i.e.~$\omega_{\phi}\ll |E_{a,b}-E_n|$ for all $n$.
For a relatively light field particle, there is no loss of generality in making this assumption, except in the case where the atomic system of interest possesses close levels of opposite parity, which will be investigated for dysprosium, ytterbium, and barium in the coming sections.

With this assumption we can present four comparatively simple formulas for the dynamic PNC amplitudes and atomic EDMs induced by the pseudoscalar interactions for both the $\g^5$ and $i\g^0\g^5$ cases presented in Eqs.~(\ref{eq:H-ig0g5}) and (\ref{eq:H-g5}):
\begin{equation}
E_{\rm PNC}^{\rm PS}(\g^5) = {\eta \omega_\phi}\sin(\omega_\phi t) K_{\rm PNC},
\label{eq:pncg5}
\end{equation}
\begin{equation}
E_{\rm PNC}^{\rm PS}(i\g^0\g^5) =
\frac{\zeta \omega_\phi}{2}\sin(\omega_\phi t) K_{\rm PNC},	
\label{eq:pncg0g5}
\end{equation}
\begin{equation}
d_{\rm EDM}^{\rm PS}(\g^5) = {-2i\eta \omega_\phi^2 \cos(\omega_\phi t)} K_{\rm EDM},
\end{equation}
and
\begin{equation}
d_{\rm EDM}^{\rm PS}(i\g^0\g^5) =  {-i\zeta \omega_\phi^2} \cos(\omega_\phi t) K_{\rm EDM}.	
\label{eq:edmg0g5}
\end{equation}

For the PV interaction presented in Eq.~(\ref{eq:H-static}), the induced PNC amplitude is given by
\begin{equation}
E_{\rm PNC}^{\rm PV}(\g^5) = b_0(t) K_{\rm PNC},
\label{eq:PVpncg5}
\end{equation}
where in the static case $b_0(t)=b_0$ 
is a constant, and in the dynamic case $b_0(t)=b_0\sin(\omega_bt)$ oscillates.
In the dynamic case, the PV $\g^5$ interaction also gives rise to an oscillating atomic EDM, given by
\begin{equation}
d_{\rm EDM}^{\rm PV}(\g^5) = {-2i b_0 \omega_b \cos(\omega_b t)} K_{\rm EDM}.
\label{eq:PVedmg5}
\end{equation}
In the above equations, we have defined $K_{\rm PNC}$ and $K_{\rm EDM}$ as
\begin{equation}
K_{\rm PNC} = \sum_{n,\i,\j} \left[ \frac{\bra{b}\v{d}_\i\ket{n}\bra{n} \g^5_\j \ket{a}}{E_a-E_n} +\frac{\bra{b}\g^5_\j\ket{n}\bra{n} \v{d}_\i \ket{a}}{E_b-E_n} \right]
\label{eq:kpnc}
\end{equation}
and 
\begin{equation}
K_{\rm EDM} = \sum_{n,\i,\j} \frac{\bra{a}\v{d}_\i\ket{n}\bra{n} \g^5_\j \ket{a}}{(E_a-E_n)^2}
\label{eq:kedm}.
\end{equation}
These quantities will henceforth be referred to as the atomic structure coefficients.

The formulas (\ref{eq:pncg5}) -- (\ref{eq:PVedmg5})  provide the connection between the atomic-structure calculations and the fundamental physics, which is necessary to extract quantitative information about the fields in question.
In deriving these equations, we made use of the relation~(\ref{eq:relation}).
Notice that the atomic structure coefficients are the same for both the $\g^5$ and $i\g^0\g^5$ cases. 
Note also that Eq.~(\ref{eq:kedm}) shows that no EDMs are induced by these fields in atomic states of zero angular momentum, since in this case the scalar operator $\g^5$ couples only intermediate states of zero angular momentum, while the vector operator $\v{d}$ cannot couple states of zero angular momentum.

For the dynamic fields, in the case where $\omega_{\phi}\sim |{E_{a,b}-E_n}|$, for a particular $n$, one has to use the complete equation (\ref{eq:pnc-general}) for the term corresponding to this $n$.
In this case, which can occur in atomic systems which possess a pair of close opposite-parity levels, there may be additional enhancement from this term.
The rest of the amplitude can be given by one of equations~(\ref{eq:pncg5}) -- (\ref{eq:PVedmg5}) with this particular term excluded.
Note that in the limit that $\omega_{\phi/b}\gg |E_{a,b}-E_n|$ for all $n$ (i.e.~a heavy field particle), the expression~(\ref{eq:pnc-general}) vanishes to lowest order.

In the nonrelativistic limit, the matrix element of the $\g^5$ operator reduces to
\begin{equation}
\bra{b}\g^5_\i\ket{a} \overset{NR}{\to} {i}(E_b-E_a)\bra{b'}\v{\sigma}_\i\cdot\v{r}_\i\ket{a'},
\label{eq:nonrel}
\end{equation}
where the {\wf}s $\ket{n'}$ are the two-component Pauli spinors (as opposed to the {\wf}s \ket{n}, which are four-component Dirac spinors).
This term scales as $1/c$; the next lowest-order corrections are of order $1/c^3$.
This means that in the nonrelativistic case, the operator $\g^5$ can be replaced with 
$i[\hat H_{NR},\v{\sigma}\cdot\v{r}]$,
 and therefore, by Eqs.~(\ref{eq:comH}) -- (\ref{eq:relation2}), the $K_{\rm PNC}$ coefficients (\ref{eq:kpnc}) vanish in the nonrelativistic limit.

In the calculations, this leads to significant cancellation between the
$\bra{b}\v{d}\ket{\delta a}$ and $\bra{\delta b}\v{d}\ket{a}$
terms in the sum~(\ref{eq:kpnc}). 
If the calculations were exact, this would eliminate the nonrelativistic part of the amplitude and leave only the relativistic corrections, constituting the correct result.
In practice, however, the cancellation leads to significant instabilities in the calculations. 
To bypass this problem, we express the $\g^5$ operator via the exact relation
\begin{equation}%
\g^5_\i = {i} [\hat H,\v{\hat\Sigma}_\i\cdot\v{r}_\i]+2\g_\i^5\hat K_\i,%
\label{eq:g5relation}%
\end{equation}%
which holds for the atomic Dirac-Coulomb Hamiltonian (\ref{eq:H-electrostat}).
Notice the similarity between the commutator term in~(\ref{eq:g5relation}) and the nonrelativistic expression~(\ref{eq:nonrel}).
Matrix elements of this commutator term between atomic states  scale as $1/c$, whereas for the $\g^5\hat K$ term they scale as $1/c^{3}$.
Here, 
$\hat K=\matr{\hat k}{0}{0}{\hat k}$, with 
$\hat k \equiv -1-\v{\sigma}\cdot\v{L}$ 
[$\hat k\Omega_\kappa=\kappa\Omega_\kappa$ for the spherical spinor $\Omega_\kappa$ with the Dirac quantum number $\kappa=(l-j)(2j+1)$], see, e.g.,~\cite{JohnsonBook2007}, $\v{L}$ and $l$ are the operator and value of the orbital angular momentum, and $j$ is the total  angular momentum of the  single-electron atomic states.
The commutator in (\ref{eq:g5relation}) cancels exactly in the amplitude, and does not contribute --- 
see Eqs.~(\ref{eq:comH}) -- (\ref{eq:relation2}).
We can, therefore, calculate the $K_{\rm PNC}$ coefficients free of large cancellation by using only the last term in (\ref{eq:g5relation}). Note that there are no such cancellations in the sum (\ref{eq:kedm}) for the $K_{\rm EDM}$ coefficients, which can be calculated directly with high numerical precision.

\subsection{Interactions with nucleons and via hadronic mechanisms}
\label{sec:nucleons}

Note that PS and PV cosmic fields can also interact with the nucleus, giving rise to nuclear anapole moments and nuclear Schiff moments, which contribute to nuclear-spin-dependent (NSD) PNC amplitudes and atomic EDMs respectively, see, e.g.,~\cite{Stadnik2014,Graham2013,Graham2011}.
In Ref.~\cite{Stadnik2014}, it was shown that an interaction of the form (\ref{eq:H-g5}) can give rise to a nuclear anapole moment (AM), a {\P}-odd, {\T}-even nuclear moment that normally arises due to parity-violating nuclear forces~\cite{Flambaum1984}.

In this work, we consider nuclear anapole moments induced by the interaction between nucleons and the static PV interaction of the form (\ref{eq:H-static}), which in the nonrelativistic limit reads
\begin{equation}
\hat W_{\rm NR} = b_0^{N}{\v{\sigma}}\cdot\v{p}/m_N,	
\label{eq:w-nuc}	
\end{equation}
where $b_0^{N}$ is the cosmic-field amplitude including the interaction strength between the cosmic field and a nucleon, and \v{\sigma}, \v{p} and $m_N$ are the spin, momentum and mass of the nucleon.
We also consider the interaction of the SME $d_{00}$  term in (\ref{eq:sme}) with nucleons.
In the nonrelativistic limit, this term leads to an interaction Hamiltonian of the form~\cite{Kostelecky1999a}
\begin{equation}
\hat W_{\rm NR} = -d_{00}^{N}\v{\sigma}\cdot\v{p}.	
\label{eq:w-nucd00}	
\end{equation}

Both interactions (\ref{eq:w-nuc}) and (\ref{eq:w-nucd00}) will contribute to the nuclear AM.
The Hamiltonian representing the NSD PNC interaction of a valence electron with the nuclear AM is given by
\begin{equation}
\hat h_{\rm AM} = \frac{G_F{K_I}}{\sqrt{2}}\frac{\v{\alpha}\cdot\v{I}}{I}\varkappa\,\rho(\v{r}),
\label{eq:h-anapole}
\end{equation}
where 
${K_I=(I+1/2)(I+1)^{-1}(-1)^{I+1/2-l_N}}$, with $l_N$ being the orbital angular momentum of the valence nucleon,
$\v{I}$ is the nuclear spin, and $\rho$ is the nuclear density~\cite{Flambaum1984} (see also~\cite{GingesRev2004}).
The dimensionless constant  
$\varkappa=\varkappa_a+\varkappa_{\rm CF}$
 quantifies the magnitude of the AM, and has contributions both from parity-violating nuclear forces, $\varkappa_a$ (the conventional AM), and from the interaction of the cosmic field with the nucleons, $\varkappa_{\rm CF}$.

From Eq.~(\ref{eq:h-anapole}), we see that the interaction of atomic electrons with the cosmic-field--induced AM has exactly the same form as their interaction with the conventional (parity-violating nuclear-force--induced) AM, the only difference being the source of the moment.
This means that no new atomic calculations are required, and a limit on the magnitude of $\varkappa_{\rm CF}$, and hence $b_0^{N}$ and $d_{00}^{N}$, can be extracted directly from existing experiments and calculations. 

The magnitude of the AM, $\varkappa_{\rm CF}$, is related to the field parameters $b_0^{N}$ and $d_{00}^{N}$ by the equation
\begin{equation}
\label{anapole_kappa_PV}
\varkappa_{\rm CF} =  \frac{2\sqrt{2} \pi  \alpha\mu_N\braket{r^2}}{G_F m_N}
(b_0^{N}-m_N d_{00}^{N}), 
\end{equation}
where $\braket{r^2}$ and $\mu_N$ are the mean-square radius and magnetic moment (in nuclear magnetons) of the valence nucleon, respectively,  and $\a\approx1/137$ is the fine-structure constant;
see Refs.~\cite{Flambaum1984,Stadnik2014} for more details.
We take $m_N = 0.94$ GeV, $\mu_p = 2.8$, $\mu_n = -1.9$, and $\braket{r^2}= (3/5) r_0^2 A^{2/3}$, where $r_0 = 1.2$ fm and $A$ is the atomic mass number.

The spin of a nucleus with an odd number of nucleons is, in general, due primarily to a single valence nucleon.
We note, however, that due to polarization of core nucleons by the valence nucleon, the nuclear spin can have contributions from both protons and neutrons. 
This means that there are contributions to the cosmic-field--induced AM coming from both protons and neutrons~\cite{StadnikNMB2014}.
In the case of cesium, which has nuclear spin $I=7/2$, this can be approximately represented via the relations
\begin{equation} 
b_0^{N} =
\frac{ b_0^{p}\mu_p\braket{\sigma^p_z} + b_0^{n}\mu_n\braket{\sigma^n_z}}{\mu_p(\braket{\sigma_z^p}+\braket{\sigma_z^n})}, 
\label{eq:b-pn}
\end{equation}
and
\begin{equation}
d_{00}^{N} =
\frac{d_{00}^{p}\mu_p\braket{\sigma^p_z} + d_{00}^{n}\mu_n\braket{\sigma_z^n}}{\mu_p(\braket{\sigma^p_z}+\braket{\sigma^n_z})},
\label{eq:d-pn}
\end{equation}
allowing us to use the measurement~\cite{Wieman1997} of the AM in cesium to constrain the interaction strengths of the considered cosmic fields with neutrons, as well as protons.
Here, the superscripts $p$ and $n$ refer to protons and neutrons, respectively, and 
$\braket{\sigma_z}$
 is the expectation value ($z$-component) of the  spin for a particular nucleon.
For thallium, which has nuclear spin $I=1/2$, this approximation is not valid, and so we only extract limits for the cosmic-field interactions with protons from the results in thallium, within the single-particle approximation; see Ref.~\cite{StadnikNMB2014} for more details.

The dynamic PS and PV  fields (\ref{eq:H-g5}) and (\ref{eq:H-static}) also induce oscillating anapole moments in atomic nuclei. 
This was considered in Ref.~\cite{Stadnik2014}.
In the case of a static PV cosmic-field--induced AM, one can immediately extract limits on the coupling of the fields with protons via the existing NSD PNC calculations and measurements in cesium~\cite{Wieman1997} and thallium~\cite{Vetter1995}. 
This is not the case for the dynamic interactions. For this reason, we consider only the static case.


The QCD axion was previously shown to give rise to oscillating {\P}- and {\T}-odd nuclear Schiff moments \cite{Graham2011,Graham2013,Stadnik2014}, which arise from {\P}- and {\T}-odd intranuclear forces and from the EDMs of constituent nucleons. 
This follows from the observation that the QCD Lagrangian contains the {\P}- and {\C\P}-violating term
\begin{equation}
\label{eq:L_QCD_CP}
\mathcal{L}_{\textrm{QCD}}^{\theta} = \theta \frac{g^2}{32\pi^2} G_a^{\mu \nu} \widetilde{G}_{a\mu \nu} , 
\end{equation}
 and that $\theta$ may be cast in the form $\theta(t) = a(t)/f_a$.
Here, $a(t)=a_0 \cos(m_a t)$ is the oscillating QCD axion field with $f_a$ the axion decay constant, $\theta$ is the dimensionless parameter that quantifies the degree of {\C\P}-violation, $G_a$ and $\widetilde{G}_a$ are the gluonic field tensor and its dual, respectively, (with color index $a$) and
$g$ is the QCD gauge coupling constant.

Here we point out that axions may also induce oscillating {\P}- and {\T}-odd effects in molecules through the generation of oscillating nuclear magnetic quadrupole moments (MQMs), which arise from {\P}- and {\T}-odd intranuclear forces and from the EDMs of constituent nucleons. 
We note that nuclear MQMs, unlike nuclear EDMs, are not screened by the atomic electrons. 
Both of these mechanisms contribute to nuclear MQMs, which are linear in $\theta$, and so recasting $\theta$ in the form $\theta(t) = a(t)/f_a$ leads to our noted inference.
Assuming that these effects are quasi-static, the approximate magnitudes of such oscillating nuclear MQMs and the effects they induce in molecules can be obtained for various cases from the numerical values in Ref.~\cite{FlambaumMQM2014} (see also 
Refs.~\cite{[][ {[Sov. Phys. JETP {\bf60}, 873 (1984)].}]Khrip1984,
[][ {[Sov. Phys. JETP {\bf44}, 25 (1976)].}]Khrip1976,
Haxton1983,Kozlov1995}) 
by the substitution 
$\theta \to a_0 \cos(m_a t) / f_a$, 
with 
$a_0/f_a \sim 4 \times 10^{-18}$ 
from consideration of the local CDM density and assuming no fine-tuning of the so-called ``misalignment angle'', see, e.g., Ref.~\cite{Stadnik2014,Graham2013,Graham2011,Ligo2009,Dine1983,Preskill1983}. 
Note that when considereing the axion field with the above assumptions, the coeficicient in Eq.~(\ref{eq:lps}) is given by
$\eta=a_0/f_a$.

\subsection{Enhancement of NSI PNC effects in diatomic molecules}

In diatomic molecules with closely spaced pairs of opposite-parity levels, only static NSD PNC effects, which are due primarily to the nuclear anapole moment, are enhanced 
\cite{[][ {[Sov. Phys. JETP {\bf48}, 608 (1978)].}]Sushkov1978,
[][ {[Sov. Phys. JETP {\bf48}, 434 (1978)].}]Labzovsky1978,Flambaum1985,Kozlov1995}
(see also \cite{DzubaReview2012}).
Static NSI PNC effects are not enhanced, since the nuclear weak charge interaction cannot mix a pair of opposite-parity rotational states. 
This may be rationalised as follows.
After averaging over the electron {\wf}, the effective operator acting on the angular variables may contain three vectors: the direction of molecular axis $\v{N}$, the electron angular momentum $\v{J}$, and the nuclear spin angular momentum $\v{I}$. 
The only {\P}-odd, {\T}-even operator that can be formed from these three vectors is proportional to 
$\v{N} \cdot (\v{J} \times \v{I})$, 
which contains the nuclear spin. 
It is also possible to form the {\P}-odd, {\T}-odd operators 
$\v{N} \cdot \v{J}$ 
and 
$\v{N} \cdot \v{I}$, 
neither of which contribute to PNC effects in the case of static interactions~\cite{DzubaReview2012}. 
However, for a time-dependent interaction of the form 
$V(t) \propto \v{N} \cdot \v{J} \cos (\omega t)$, 
there arises an additional term in the perturbed molecular {\wf} that is shifted in phase by $\pi / 2$ radians compared to the original term, which is the only (real) term present in the case of a static interaction of the form $\v{N} \cdot \v{J}$ --- compare with Eq.~(\ref{eq:cnt}).  
Hence there may be enhancement of both NSI and NSD PNC effects in diatomic molecules possessing close pairs of opposite-parity levels in the presence of time-dependent interactions.

\section{Methods for atomic structure calculations}\label{sec:calcs}

We examine a number of different systems, and use different computational methods for the 
{\sl ab initio}~relativistic calculations.
We outline these briefly and refer the reader to the relevant sources for more detailed information.

\subsection{Single-valence electron systems}

For atoms and ions with one valence electron above a closed-shell core, we employ the correlation potential method~\cite{Dzuba1987jpbRPA,DzubaCPM1989plaPNC,*DzubaCPM1989plaEn,
*DzubaCPM1989plaE1}.
We start from the mean-field Dirac-Fock approximation with a $V^{N-1}$ potential and  include dominating electron correlation effects. 
The correlation potential, $\hat \Sigma_1$, which includes a summation of the series of dominating diagrams, is calculated to all orders of many-body perturbation theory using relativistic Hartree-Fock Green's functions and the Feynman-diagram technique~\cite{DzubaCPM1989plaPNC,DzubaCPM1989plaEn,
*DzubaCPM1989plaE1}.  
We also calculate the correlation potential to only second order ($\hat \Sigma^{(2)}$), for use when the all-order method is not appropriate, and as a test of the accuracy~\cite{Dzuba1987jpbRPA}.
The correlation potential $\hat \Sigma_1$  is then used to construct the set of so-called Brueckner orbitals (BOs) for the valence electron, which  are found by solving the Hartree-Fock-like equations including the operator $\hat \Sigma$:   
\begin{equation}
 (\hat H_0 +\hat \Sigma_1 - \en_n)\psi_n^{(\rm BO)}=0.			%
\label{eq:BO}
\end{equation}
Here, 
\begin{equation}
 \hat H_0 =   \v{\alpha}\cdot\v{p} + m_e(\g^0 -1) - V^{\rm nuc}+U^{\rm HF}
\label{eq:h0}
\end{equation}
is the relativistic Hartree-Fock (RHF) Hamiltonian 
with nuclear potential, $V^{\rm nuc}$, and Hartree-Fock potential, $U^{\rm HF}$;
$\en_n$ is the single-particle energy corresponding to the Bruckner orbital $\psi_n^{(\rm BO)}$,  and the index $n$ denotes valence states.
Core polarization and the PNC and $E1$ interactions are included via the time-dependent Hartree-Fock (TDHF) method~\cite{Dzuba1987jpbRPA,DzubaCPM1989plaPNC,*DzubaCPM1989plaEn,
*DzubaCPM1989plaE1}, which is sometimes also referred to as the RPA (random phase approximation) method.

To calculate the core-polarization corrections, we write the single-electron {\wf} in an external PNC and $E1$ field using the TDHF method as 
\begin{equation}
\psi = \psi_0 + \delta\psi +X e^{-i\omega t}+Y e^{i\omega t},
\end{equation}
where $\psi_0$ is the unperturbed state, $\delta\psi$ is the
correction due to the cosmic-field--induced PNC interaction acting alone,
$X$ and $Y$ are corrections due to the $E1$ interaction acting alone,
and $\omega=|E_a-E_b|$ is the frequency of the PNC transition ($\omega=0$ for EDMs). 
 These corrections are found by solving the system of TDHF equations self-consistently for the core: 
\begin{align}
(\hat H_0 - \en_c)\delta\psi_c &= - (\hat h_{\g^5} + \delta \hat
V_{\g^5})\psi_{0c}, \nonumber \\ 
(\hat H_0 - \en_c-\omega)X_c &= - ( \v{d} + \delta \hat
V_{E1})\psi_{0c},  \label{eq:RPA} \\  
(\hat H_0 - \en_c+\omega)Y_c &= - ( \v{d}^{\dagger} + \delta \hat
V_{E1}^{\dagger})\psi_{0c},    \notag  
\end{align}
where the index $c$ denotes core states,  and
$\delta \hat V_{\g^5}$ and $\delta \hat V_{E1}$ are corrections to the core potential arising from the PNC and $E1$ interactions respectively.
Note that in the equations~(\ref{eq:RPA}), we have neglected the contribution from $\omega_{\phi}$, i.e.~we have assumed that $\omega_{\phi}\ll |E_{\rm core} - E_{a,b}|$.
The core excitation energy is very large, so this should be valid in all cases.

The PNC and EDM atomic structure coefficients  (\ref{eq:kpnc}) and (\ref{eq:kedm}) can then be  calculated using single-particle energies and {\wf}s, with the operators $\v{d}_\i$ and $\g^5_\i$ replaced by the effective single-particle operators including the core-polarization corrections: $\sum_\i\v{d}_\i\to\v{d}+\delta\hat V_{E1}$, $\sum_\i\g^5_\i\to\g^5+\delta \hat V_{\g^5}$. 
This is how we calculate the $K_{\rm EDM}$ values, however, for the $K_{\rm PNC}$ values we use a slightly different method due to the instabilities caused by the large cancellation discussed previously.

By expressing the second term on the right-hand side of (\ref{eq:g5relation}) as $2\g^5\hat K=-2\g^0\g^5(\g^0\hat K)$, and noting that single-particle states are eigenstates of $\g^0\hat K$ (with eigenvalue $\kappa$), we can use Eq.~(\ref{eq:relation}) 
to express the PNC (single-particle) matrix elements as
\begin{equation}
\bra{\psi_n} 2\g^5\hat K \ket{\psi_a} = \frac{-\kappa_a}{2m_e}(\en_n-\en_a) \bra{\psi_n}\g^5  \ket{\psi_a} .
\label{eq:spme}
\end{equation}
Upon substitution into the summation for $K_{\rm PNC}$, we can invoke the closure relation and the amplitude for single-particle states reduces to
\begin{equation}
K_{\rm PNC} = \frac{1}{m_e}(\kappa_b+\kappa_a)\bra{\psi_b}\g^5(\v{d}+\delta V_{E1})\ket{\psi_a},
\label{eq:newkpnc}
\end{equation}
where we have neglected the core polarization due to the $2\g^5\hat K$ operator, since it is highly suppressed. 
This expression requires no summation over intermediate states, does not contain significant cancellation, and can be calculated with relatively high accuracy.
We include correlations by using the BOs $\psi_a$ and $\psi_b$ for the valence states $a$ and $b$ in Eq.~(\ref{eq:newkpnc}).

For the $K_{\rm EDM}$ coefficients, the first term on the right-hand side of Eq.~(\ref{eq:g5relation}) does not cancel.
 In fact, this term dominates the amplitude [since it leads almost directly to the nonrelativistic approximation~(\ref{eq:nonrel})]
and scales as $1/c$, whereas the second term scales as $1/c^3$.
Inserting {${\g^5\approx i [\hat H,\hat{\v{\Sigma}}\cdot\v{r}]}$}
into (\ref{eq:kedm}),  we see that the $K_{\rm EDM}$ coefficients for ${}^2S_{1/2}$ states are approximately proportional to the static dipole polarizability, with corrections on the order of $(1/c)^3$.
The constant of proportionality is determined by Eq.~(\ref{eq:g5relation}) and the angular integrals~\cite{Varshalovich1988}:
\begin{align}
K_{\rm EDM}(z) 
&\simeq -{i} \sum_{n} \frac{\bra{a}{d}_z\ket{n}\bra{n} \v{\hat\Sigma}\cdot\v{r} \ket{a}}{E_a-E_n} \notag\\
&\approx \frac{i}{2e} \alpha_0,
\label{eq:edma0}
\end{align}
where the scalar electric dipole polarizability, $\alpha_0$, is given by
\begin{equation}
\alpha_0 = -\frac{2e^2}{3(2J_a+1)}\sum_{n} \frac{\left|\bra{a}|r_z|\ket{n}\right|^2}{E_a-E_n},
\end{equation}
where $\bra{a}|r_z|\ket{n}$ is the $z$ component of the reduced matrix element of the \v{r} operator.
[Equation (\ref{eq:edma0}) relies on the fact that the radial integrals and energies depend only on the $n,l$ quantum numbers, and not on $j$, in the nonrelativistic limit.]
This can be used as an independent test of the calculations.
Rougher (and far less accurate) relations can also be derived for other states, e.g.~the $^2P_{1/2}$ ground state of thallium, which are useful for order-of-magnitude estimates.

Note that in the methods described above we have not included the core polarization contribution that comes from the simultaneous action of the $E1$ and PNC fields, the so-called ``double core polarization'', see, e.g.,~\cite{DzubaCPM1989plaPNC,*DzubaCPM1989plaEn,
*DzubaCPM1989plaE1,RobertsDCP2013}. 
Core polarization amounts to only a small correction to the quantities considered in this work, so the even smaller double core polarization can be safely neglected in most cases.
In the case of thallium, however, where the single-particle approach is less valid, this may have a significant impact on the accuracy.

\subsection{Two valence electron atoms}

We treat ytterbium and barium as systems with two valence electrons above a closed shell core, and follow closely the methods employed recently~\cite{Dzuba2011Yb,RobertsCL2014} to calculate conventional PNC effects in these atoms.
Starting from the RHF method with the potential $U^{\rm HF}$ created by the $N-2$ electrons of the closed-shell core \cite{DzubaVN-M2005}, where $N$ is the total number of electrons,
 we make use of the combined configuration interaction (CI) and many-body perturbation theory (MBPT) method developed in Ref.~\cite{DzubaCIMBPT1996}. 
Interactions with external fields and core polarization are taken into account using the TDHF method as above.
For more detail on this method, see also Refs.~\cite{Dzuba2011Yb,DzubaYbPol2010,DzubaCIMBPT1998}.

The effective CI+MBPT Hamiltonian for the system of two valence electrons has the form:
\begin{equation}
\hat H^{\rm eff} = \hat h_1(\v{r}_1)+\hat h_1(\v{r}_2) + \hat h_2(\v{r}_1,\v{r}_2),
\label{eq:hci}
\end{equation}
where $\hat h_1$ is the single-electron part of the RHF Hamiltonian,
\begin{equation}
 \hat h_1 =   \v{\alpha}\cdot\v{ p} + m_e(\g^0 -1) - V^{\rm nuc}+U^{\rm HF} + \hat\Sigma_1,
\label{eq:h1}
\end{equation}
and $\hat h_2$ is the two-electron part,
\begin{equation}
 \hat h_2(\v{r}_1,\v{r}_2) = \frac{e^2}{r_{12}} + \hat\Sigma_2(\v{r}_1,\v{r}_2).
\label{eq:h2}
\end{equation}
The additional terms, $\hat\Sigma$, are the correlation potentials, which are used to take into account core-valence correlations (see Refs.~\cite{DzubaCIMBPT1996,DzubaCIMBPT1998} for details).
The single electron correlation potential, $\hat\Sigma_1$, is the same potential as described above (here we use only the second-order correlation potential, $\hat\Sigma^{(2)}$), and represents the interaction of a single valence electron with the atomic core.
The two-electron operator, $\hat\Sigma_2$, represents the screening of the valence-valence Coulomb interaction by the core electrons.

We also introduce a scaling parameter: $\hat\Sigma_1\to\lambda_\kappa\hat\Sigma_1$ in (\ref{eq:h1}), where $\lambda_\kappa$ can take different values for different values of $\kappa$ ($s_{1/2}$, $p_{1/2}$ etc.)~and $\lambda_\kappa\approx1$.
The scaling parameters serve two purposes. 
Firstly, since the single-particle energies in Eq.~(\ref{eq:spme}) are relatively sensitive to $\lambda_\kappa$, whereas the radial integrals are comparatively insensitive, we can use this as a test of the stability of the calculations. 
We do this and find satisfactory stability for both the matrix elements and the overall PNC amplitudes.
Secondly, in the case of the PNC transition in ytterbium, a system that possesses a pair of relatively close levels of opposite parity, we can use the scaling parameters to fit the important energy differences to the experimental energies. This is important, since even modest errors in individual energy levels may lead to an error of orders-of-magnitude in an energy interval when it is particularly small. 
See Ref.~\cite{RobertsCL2014} for a detailed discussion on this point.

The matrix elements are then computed from the sum of the single-particle contributions.
For the single-particle contributions, we use Eq.~(\ref{eq:spme}), which removes all significant cancellation into a small factor $\sim1/c^3$ [two factors of $c$ come from the coefficient $m_e$ in (\ref{eq:spme}), the third comes from the lower (small) component of the Dirac radial {\wf}].

Note that we can also use Eq.~(\ref{eq:nonrel}) to approximately express Eq.~(\ref{eq:spme}) as 
\begin{equation}
\bra{\psi_n} 2\g^5\hat K \ket{\psi_a} \approx \frac{-i\kappa_a}{2m_e}(\en_n-\en_a)^2 \bra{\psi_n'}
\v{\sigma}\cdot\v{r}  \ket{\psi_a'},
\label{eq:se-nonrel}
\end{equation}
where the corrections are of order $1/c^3$.
Equations (\ref{eq:spme}) and (\ref{eq:se-nonrel}) have very different radial integrals;
as such, performing the calculations using both these equations serves as a good numerical test of our method.
We find good agreement between both the matrix elements and the amplitudes calculated using Eqs.~(\ref{eq:spme}) and (\ref{eq:se-nonrel}). 
This is important, since it justifies neglect of core polarization due to the $2\g^5\hat K$ operator.

\subsection{Dysprosium}

The feature of dysprosium that makes it a particularly interesting system for the study of atomic PNC is the existence of two nearly degenerate states of opposite parity and the same total angular momentum, $J=10$, at $E = 19797.96$ cm$^{-1}$.
We use the notation $A$ for the even-parity state and notation $B$ for the odd-parity state, following Ref.~\cite{Nguyen1997}. 
The PNC experiment in dysprosium is different to those done, for example, in cesium, and it is the quantity $\bra{A}\gamma^5\ket{B}$ that is of most interest.
This is because, in dysprosium, the mixing of the opposite parity $A$ and $B$ states is observed directly, whereas in in the other experiments it is transitions between states of the same parity that are observed~\cite{Nguyen1997} (the parity-violating part of these transitions is enabled by a mixing of many opposite-parity states).

The method we use for the calculations in dysprosium follows almost exactly previous calculations of conventional PNC effects in this system~\cite{DzubaDy2010}, with the only exception being the interchange of the operator of the  electron-nucleus weak interaction (\ref{eq:hqw}) with those for the parity-violating interactions with cosmic fields,~(\ref{eq:H-g5}) and (\ref{eq:H-static}). 
We use the particular CI method described in much greater detail in Ref.~\cite{DzubaVN2008}.
To construct the single-electron orbitals, we use a $V^N$ potential, where $N=66$ is the total number of electrons.

A different $V^N$ Hartree-Fock potential  is used for each different configuration, then the valence states found in the Hartree-Fock calculations are used as basis states for the CI calculations. 
This helps account for the fact that single-electron states actually depend on the configurations.  
While it is possible to account for this dependence within the CI calculations, it requires a complete set of single-electron states. 
These would then be used to construct the many-electron basis states by redistributing the valence electrons over the single-electron basis states.
Then the actual many-electron states are found by diagonalizing the matrix of the effective CI Hamiltonian~\cite{DzubaVN-M2005}. 
This approach works well in the case of a few valence electrons, e.g.~neutral barium and radium as discussed above.
 However, for the twelve valence electrons of dysprosium, it would lead to a matrix of enormous size making it practically impossible to saturate the basis with limited computing resources. 
The results with an unsaturated basis are unstable and strongly depend on where the basis is truncated. 
Therefore, it is preferable to account for the differences in the configurations at the Hartree-Fock, rather than the CI, stage of the calculations.


After the self-consistent Hartree-Fock procedure is done for each necessary configuration, the effective CI Hamiltonian for the valence states of dysprosium, with $M=12$ valence electrons, is expressed as
\begin{equation}
\hat H^{\rm eff} = \sum_{\i=1}^{M} \hat h_1(\v{r}_\i)+  \sum_{\j<\i} \frac{e^2}{r_{\i\j}},
\label{eq:hci-dy}
\end{equation}
where
\begin{equation}
 \hat h_1 =   \v{\alpha}\cdot\v{p} + m_e(\g^0 -1) - V^{\rm nuc}+U^{\rm HF} + \delta V_{\rm p}.
\label{eq:h1-dy}
\end{equation}
Here $U^{\rm HF}$ is the Hartee-Fock potential due to the $N-M$ core electrons.
We do not use the {\sl ab initio} correlation potential as described above, instead it is the term $\delta V_{\rm p}$ in Eq.~(\ref{eq:h1-dy}) that simulates the effect of valence-core correlations. It is known as the polarization potential, and has the form
\begin{equation}
\delta V_{\rm p} = -\frac{\alpha_{\rm p}}{2(r^4+a_0^4)},
\label{eq:pol-pot}
\end{equation}
where $\alpha_{\rm p}$ quantifies the polarization of the core, and $a_0$ is a cut-off parameter, for which we use the Bohr radius.
The term $\alpha_{\rm p}$  is treated as a parameter and is scaled to reproduce the correct experimental energies. 
The effect that adding or removing basis configurations, and making small changes in the values $\alpha_{\rm p}$, has on the amplitude is a good way to test the accuracy of the calculations.

Since the states of interest in dysprosium are practically degenerate, the commutator term in Eq.~(\ref{eq:g5relation}) does not contribute to the matrix element.
We therefore calculate the matrix elements of the PNC interaction directly from the single-particle contributions using Eq.~(\ref{eq:spme}).
We use the same configurations and values for $\alpha_{\rm p}~(\approx0.4$ a.u.) as in Ref.~\cite{DzubaDy2010}.

\section{Results and Discussion}\label{sec:results}

  \begin{table}%
    \centering%
    \caption{ Calculations of the PNC and EDM atomic structure coefficients [$j_z=\min(j_a,j_b)$] for several atomic systems. 
Valid in the case that $\omega_{\phi}\ll |E_{a,b}-E_n|$. Values are presented in atomic units.} 
\begin{ruledtabular}%
  \begin{tabular}{l l D{,}{}{3.8} c D{.}{.}{3.7}} 
        &\multicolumn{2}{c}{ PNC}      & \multicolumn{2}{c}{EDM} \\
\cline{2-3}\cline{4-5}
        & \multicolumn{1}{c}{Transition} & \multicolumn{1}{c}{$K_{\rm PNC}\,(i10^{-6})$} & \multicolumn{1}{c}{State} & \multicolumn{1}{c}{$K_{\rm EDM}$} \\
\hline
H & $1s$-$2s$					& 0,.1447(2)& $1s$		& 0.0164(1)\\  
Li & $2s$-$3s$ 					& 0,.219(3)	& $2s$ 		& 0.60(1) \\ 
Na & $3s$-$4s$ 				& 0,.224(4)	& $3s$ 		& 0.61(1) \\ 
K & $4s$-$5s$ 					& 0,.242(4)	& $4s$ 		& 1.09(5) \\ 
 & $4s$-$3d_{3/2}$ 			& -0,.307(6)&  			&  		\\ 
Cu	&							&			& $4s$		& 0.16(3)\tablenotemark[1]	\\
Rb & $5s$-$6s$ 				& 0,.247(5)	& $5s$ 		& 1.22(8) \\ 
Ag	&							&			& $5s$		& 0.17(5)\tablenotemark[1]	\\
 & $5s$-$4d_{3/2}$ 			& -0,.30(1)	&  			&  		\\ 
Cs & $6s$-$7s$					& 0,.256(5)	& $6s$ 		& 1.6(2) \\ 
 & $6s$-$5d_{3/2}$ 			& -0,.22(3)	&  			&  		\\ 
Ba&${}^1S_0$-${}^3D_1$		& -0,.5(1)	&			&		\\ 
Ba$^+$ & $6s$-$5d_{3/2}$ 	& -0,.02(1)	&  			&  		\\ 
Yb & ${}^1S_0$-${}^3D_1$ 	& -8,(2) 	&			&		\\  
Au	&							&			& $6s$		& 0.12(4)\tablenotemark[1]	\\
Tl & $6p_{1/2}$-$6p_{3/2}$ 	& 0,.22(5) 	&$6p_{1/2}$&   0.2(1)\\   
Fr & $7s$-$8s$ 					&  0,.253(6)	& $7s$ 		& 1.3(2) \\   
 & $7s$-$6d_{3/2}$ 			& -0,.25(3)	&  			&  		\\ 
Ra$^+$ & $7s$-$6d_{3/2}$ 	& -0,.08(3) 	&  			&		\\
  \end{tabular}%
\end{ruledtabular}%
\tablenotetext[1]{From polarizability calculations~\cite{Schwerdtfeger1994,*Neogrady1997,*Roos2005}.}
    \label{tab:calculations}%
  \end{table}

\subsection{Values and accuracy of the atomic structure coefficients}

Results of our calculations for the atomic structure coefficients $K_{\rm PNC}$ and $K_{\rm EDM}$ [defined in equations (\ref{eq:pncg5}) through (\ref{eq:kedm})] are presented in Table~\ref{tab:calculations}.
We present $z$-components, with $j_z={\rm min}(j_a,j_b)$.

In order to estimate the uncertainty, we calculate the values $K_{\rm PNC}$  without including any correlations, including correlations to second-order ($\Sigma^{(2)}$), and including correlations to all-orders (see Sec.~\ref{sec:calcs}).
We take the all-order results as the midpoint, and estimate the uncertainty as the difference between this and the pure Hartree-Fock (no correlations) calculations. 
The second-order results are used as an extra test; the deviation of the second-order results from the all-order ones is significantly smaller than the assumed uncertainty.
We also examine the effect that including core polarization has on the amplitudes and note that its effect is also smaller than the assumed uncertainty.

\begin{figure}%
\begin{center}%
\includegraphics[width=4.5cm]{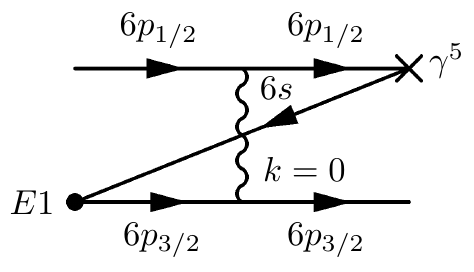}%
\caption{Example Feynman-Goldstone diagram for the contribution to the cosmic PNC transition (\ref{eq:kpnc}) in thallium arising from the double core polarization by the PNC cosmic-field (cross) and the electric-dipole (dot) interaction. 
The 6s state is treated as a state in the core. The wavy line is the Coulomb interaction with multipolarity  $k$.}%
\label{fig:tl-dcp}%
\end{center}%
\end{figure}

Note that we treat thallium here as a single valence electron system, where the $6s^2$ electrons are treated as core states.
In order for this treatment of thallium to yield accurate results one needs to take into account many higher-order correlation corrections, such as ladder-diagrams~\cite{DzubaLadder2008}.
In particular, the double core polarization may give a significant contribution in this approximation, see Fig.~\ref{fig:tl-dcp}.
Therefore, for the Tl $K_{\rm PNC}$ we use only the second-order correlation potential, and the uncertainty is taken as the size of these correlation effects. 
The uncertainty attributed to thallium takes into account the omitted core-polarization effects.
An alternative method for calculations in thallium is to treat it as a three-valence-electron system, and use the CI+MBPT method, see, e.g.,~\cite{KozlovTlEDM2012}. 
In this approach, the double core polarization is taken into account automatically.
The trivalent CI+MBPT method is significantly more computationally demanding than the methods we employ in this work, and is not necessary at the currently desired level of accuracy; more complete calculations can be performed when further experimental work in this area is undertaken.

For hydrogen, we perform the calculations both using exact Dirac-Coulomb {\wf}s and numerical {\wf}s including finite-nuclear-size effects. The difference between these two approaches is negligible at the desired accuracy.
The $\bra{2s}\g^5\ket{2p}$ matrix element is almost identically zero numerically (without including radiative corrections). This means that despite being a seemingly good candidate for a Dy-type stark-interference experiment, where the PNC matrix element is measured directly (see~\cite{Nguyen1997}), hydrogen is unlikely to yield informative results in this case.
The uncertainty estimates in the hydrogen $1s$-$2s$ $K_{\rm EDM}$ value comes mainly from a truncation of the basis used for the summation, and the uncertainty for the $1s$ $K_{\rm PNC}$ value reflects the omission of QED effects, which become important at this scale ($\sim1/c^3$).

In the case of atomic EDMs, there is no cancellation as for the $K_{\rm PNC}$ values, and these magnitudes are comparatively stable.
The accuracy of these calculations is expected to be relatively high, with the dominating uncertainty coming from the inclusion of electron correlations.
We take as an estimate of the uncertainty the difference between the calculations performed with the second-order and the all-order correlation potential.
As noted above, the expression for the EDM atomic structure coefficients (\ref{eq:kedm})
can be reduced to a form very similar to that of the electric dipole scalar polarizability~(\ref{eq:edma0}).
We use this fact as a test of our calculations and find excellent agreement using published polarizability values; better than 1\% for lithium and sodium, and better than 5\% for most other atoms, see, e.g.,~\cite{Schwerdtfeger2014c}.  
The decline in agreement for the higher $Z$ systems is due to the larger role of relativistic effects here, since Eq.~(\ref{eq:edma0}) is a nonrelativistic approximation.

From the results in Table~\ref{tab:calculations}, we see that the magnitudes of PNC amplitudes in general increase with increasing atomic mass.
This can be understood as a relativistic effect, since the amplitude vanishes in the nonrelativistic limit.
However, we note that the magnitudes increase considerably more slowly with $Z$ than the $Z^3$ dependence of conventional NSI PNC effects induced by $Z^0$-boson exchange between atomic electrons and nucleons \cite{Bouchiat1974a,Bu2008}. 
This means that light atoms may also be suitable candidates for searches of pseudoscalar and pseudovector cosmic-field--induced effects.

Since the considered interaction is one with an external cosmic field, as opposed to a nuclear-sourced field as in the case of conventional atomic PNC, the amplitudes are not necessarily restricted by the value of the {\wf}s on the nucleus.
In conventional PNC, this has the effect of greatly suppressing contributions from higher orbital angular momentum ($l$) states, in which electrons do not spend as much time near the nucleus. 
This limits the magnitude of the PNC effect in many transitions, such as the $A$-$B$ matrix element in dysprosium, that have otherwise ideal conditions (high nuclear charge $Z$, very close opposite-parity levels). 
Such restrictions were noted very early, see, e.g.,~\cite{Khriplovich1991}.
In the cosmic-field--induced PNC effect, however, this restriction does not apply.


  \begin{table*}%
    \centering%
    \caption{Matrix elements of the $2\g^5\hat K$ operator for Ba, Ra, Dy, and Yb between 
			nearly-degenerate opposite-parity levels.} 
\begin{ruledtabular}%
  \begin{tabular}{lrrdd}
 & \multicolumn{1}{c}{A} & \multicolumn{1}{c}{B} & \multicolumn{1}{c}{$\Delta E_{BA}$ (cm$^{-1}$)~\cite{NISTc}} 
&\multicolumn{1}{c}{ $\bra{B} 2\g^5\hat K \ket{A}$ ($i$ a.u.)\tablenotemark[1]} \\ 
\hline
Ba & $5d^2\,({}^1D_2)$ & $5d6p\,({}^1D^o_2)$ & -12.34 & 0.3(1)\E{-9}\\ 
Dy & $4f^{10}5d6s\,(J=10)$ & $4f^{9}5d^26s\,(J=10)$ &  & 0.7(2)\E{-8}  \\ 
Yb &  $5d6s\,({}^3D_1)$  &$6s6p\,({}^1P^o_1)$&-579.12  &0.29(6)\E{-8} \\
  \end{tabular}%
\end{ruledtabular}%
\tablenotetext[1]{For ease of comparison with the literature, note that $0.7\E{-8}\mathrm{~a.u.}=50$ MHz.}
    \label{tab:Dy}%
  \end{table*}%

For the dynamic interactions, the results presented in Table~\ref{tab:calculations} are valid only in the case that
{$\omega_{\phi}\ll |E_{a,b}-E_n|$}. 
As stated above, this should generally not be a problem, except for when there exists a pair of close opposite parity levels in the summation~(\ref{eq:pnc-general}). 
Such a pair of close levels appears in barium, dysprosium, and ytterbium.
In Table~\ref{tab:Dy}, we present calculations of the $2\g^5\hat K$ matrix element between states that correspond to close levels of opposite parity in these atoms.

For dysprosium, it is actually the quantity $\bra{B} \gamma^5 \ket{A}$, as opposed to the PNC amplitude $E_{\rm PNC}$, that is directly of interest, since the transitions between $B$ and $A$ are directly measured in the dysprosium experiments.
To determine the uncertainty in this quantity, we examine the effect of removing configuration states from the basis.
Note that in the conventional PNC case, the $\bra{A}\hat h_{Q_W}\ket{B}$ matrix element is highly dependent on the configurations used~\cite{DzubaDy2010}.
We perform the calculations including only the leading two configurations for each state, as well as including all twelve of the configurations considered in~\cite{DzubaDy2010}, and many combinations in between.
We find, in fact, that this makes little difference to the final amplitude, meaning it is quite stable.
We take the uncertainty in this value to cover the range of values obtained between using only the leading two configurations for each state and using all twelve considered basis configurations.
Despite making relatively large changes to the energies, modest modifications to $\alpha_{\rm p}$ make only small changes to the amplitude; smaller than the assumed level of accuracy.

\subsection{Limits on the interactions of a pseudovector cosmic field}

For the static case, the PV interaction will manifest itself as a small addition to the PNC amplitude of a transition between two states of the same nominal parity.
Therefore, by combining the results of the conventional ($Q_W$ induced) PNC experiments and calculations with the calculations of the cosmic-field--induced PNC amplitude [given by Eq.~(\ref{eq:PVpncg5}) and Table~\ref{tab:calculations}], it is possible to extract limits on the values of the PV cosmic-field coupling constants $b_0$.
We present these limits in Table~\ref{tab:limit}.

  \begin{table*}%
    \centering%
    \caption{  Comparison of calculated and observed PNC amplitudes in Cs, Tl and Yb, and the relevant weak matrix element in Dy, and extraction of limits on the electron--cosmic-field interaction parameter, $b_0^e$. } 
\begin{ruledtabular}%
\begin{tabular}{l l D{,}{}{3.2} l D{,}{}{3.2} l r}
 &  & \multicolumn{4}{c}{$E_{\rm PNC}^{Q_W}$ ~($i 10^{-11}$  a.u.)}  &  \\ 
\cline{3-6}
\multicolumn{2}{c}{Transition}   & \multicolumn{2}{c}{Experiment}   & \multicolumn{2}{c}{Theory} & \multicolumn{1}{c}{ $|b_0^e|$ limit (GeV)}\\ 
\hline
Cs 	& $6s$ -- $7s$ 
		& 0,.8353(29) 	& \cite{Wieman1997} 	& 0,.8428(38)&\cite{OurCsPNC2012} & 2\E{-14} \\ 
Tl 	& $6p_{1/2}$ -- $6p_{3/2}$ 
		& 24,.8(2) 	& \cite{Vetter1995} 		& 25,.6(7)&\cite{Kozlov2001} & 2\E{-12} \\ 
Yb	& ${}^1S_0$ -- ${}^3D_1$ 
		& 87,(14) 	& \cite{Tsigutkin2009,*Tsigutkin2010} & 110,(14)&\cite{Dzuba2011Yb} & 2\E{-12} \\ 
\hline
\hline 		
&&&&&&\\ 
 &  & \multicolumn{4}{c}{$\bra{A}\hat h_{Q_W}\ket{B}$ ~($i 10^{-16}$  a.u.)\tablenotemark[1]}  &  \\ 
\cline{3-6}
\multicolumn{2}{c}{}   & \multicolumn{2}{c}{Experiment}   & \multicolumn{2}{c}{Theory} & \multicolumn{1}{c}{ $|b_0^e|$ limit (GeV)}\\ 
\hline
Dy &  & 3,.5(4.5) & \cite{Nguyen1997} & 6,(6)&\cite{DzubaDy2010} & 7\E{-15} \\     
  \end{tabular}%
\end{ruledtabular}%
\tablenotetext[1]{$3.5\E{-16}\mathrm{~a.u.}=2.3$ Hz; $6(6)\E{-16}\mathrm{~a.u.}=4$ Hz.}
    \label{tab:limit}%
  \end{table*}

The most stringent limit comes from the results in dysprosium.
This is due mainly to the significantly low absolute uncertainty in both the theoretical and experimental limits on the $\hat h_{Q_W}$ matrix element.

We have used the available NSD PNC measurements for cesium and thallium to extract limits on the constants $b_0^p$ and $b_0^n$ that quantify the interaction strength of a PV cosmic field (\ref{eq:w-nuc}) with protons and neutrons, respectively.
We also use these measurements to constrain the constants 
$d_{00}^p$ and $d_{00}^n$ that appear in (\ref{eq:w-nucd00}), which quantify the interaction strengths of protons and neutrons with the SME $d_{\mu\nu}$ tensor field (\ref{eq:smehG}). 
We present these limits in Table~\ref{tab:kappa}.
In extracting the limits, we have taken the values of the conventional (nuclear-forced induced) AM as $\varkappa_a=0.19$ and assumed a 30\% uncertainty for the nuclear theory for cesium, and $\varkappa_a=0.17$ with 60\% uncertainty for thallium, see, e.g., Ref.~\cite{GingesRev2004}.
The nuclear spin in both cesium and thallium is primarily due to the valence protons.
For thallium, we use a single-particle picture and therefore extract limits for the proton only.
For cesium, we use Eqs.~(\ref{eq:b-pn}) and (\ref{eq:d-pn}), along with values for $\braket{{\sigma}^p_z}$ and $\braket{{\sigma}^n_z}$, from Ref.~\cite{StadnikNMB2014}, to determine the proton and neutron limits. 
The differences between the $b_0^p$ limits for cesium presented in Table~\ref{tab:kappa} and those of Ref.~\cite{RobertsCosmic2014} is that in \cite{RobertsCosmic2014} we used the single-particle approximation.

  \begin{table*}%
    \centering%
    \caption{ Theoretical and observed values for the nuclear AM constant $\varkappa_a$ for Cs and Tl, and the extracted limits on the proton-- and neutron--cosmic-field interaction parameters, $b_0^{p,n}$ and $d_{00}^{p,n}$. } 
\begin{ruledtabular}%
\begin{tabular}{rllllllll}
 & \multicolumn{4}{c}{$\varkappa_a$}    
& \multicolumn{2}{c}{$b_0$ limits (GeV)}   
& \multicolumn{2}{c}{$d_{00}$ limits}\\ 
\cline{2-5}\cline{6-7}\cline{8-9}
\smallspace
 & \multicolumn{2}{c}{Observed}   & \multicolumn{2}{c}{Theory}  
  & \multicolumn{1}{c}{$|b_0^p|$} 
  & \multicolumn{1}{c}{$|b_0^n|$}  
 & \multicolumn{1}{c}{$|d_{00}^p|$} 
 & \multicolumn{1}{c}{$|d_{00}^n|$} \\ 
\hline
\smallspace
$^{133}$Cs 		& \phantom{$-$}0.364(62) & \cite{Wieman1997,FlambaumAnM1997} & 0.15 --- 0.23 &\cite{Dmitriev1997,Dmitriev2000,*Haxton2001c,*Haxton2002} 		 & 4\E{-8} &2\E{-7}&5\E{-8}&2\E{-7}\\ 
$^{203,205}$Tl 	& $-$0.22(30) & \cite{Vetter1995,Khriplovich1995} & 0.10 --- 0.24 & \cite{Dmitriev2000,*Haxton2001c,*Haxton2002} & 8\E{-8} &&9\E{-8}&\\ 
\end{tabular}
\end{ruledtabular}%
    \label{tab:kappa}%
  \end{table*}%

These field-nucleon coupling limits are to be compared with the field-electron coupling limits obtained from PNC amplitude measurements and from direct determination of weak interaction matrix elements, which are tabulated in Table~\ref{tab:limit}. The latter limits are by far the more stringent.  
Note that ongoing AM measurements with Fr, Yb, and BaF will also lead to limits on PV cosmic-field couplings to protons and neutrons~\cite{Tsigutkin2009,*Tsigutkin2010,Aubin2013,DeMille2008,Cahn2014}.

\subsection{Experimental accessibility of dynamic effects}

After the first observation in bismuth~\cite{[][{ [Pis'ma Zh. Eksp. Teor. Fiz {\bf27}, 379 (1978)].}]Barkov1978}, conventional atomic PNC effects have since been observed in lead, cesium,  thallium, and ytterbium,  
see Refs.~\cite{FortsonPb1993,Meekhof1995,Wieman1997,Lintz2007,
Vetter1995,Tsigutkin2009,*Tsigutkin2010} 
and references within.
Atomic PNC experiments have also been proposed for the here considered single-valence systems 
 francium, rubidium, Ba$^+$, and Ra$^+$,  see,
 e.g.,~\cite{Shabaev2005,Aubin2013,Safronova2000,OurRb2012,Fortson1993,
			NunezPortela2013,Pal2009,RobertsActinides2013,*Robertssd2014},
as well as barium, radium, and other heavy elements~\cite{RobertsCL2014}, and are ongoing for dysprosium~\cite{Nguyen2000,Weber2013a}.

Of the atoms considered here, EDM measurements have been performed using the rubidium~\cite{Ensberg1967}, cesium~\cite{Murthy1989}, and thallium~\cite{Regan2002} atoms.
They have also been performed using mercury~\cite{Griffith2009,Swallows2013}, xenon and helium~\cite{Rosenberry2001}, and
the meta-stable ${}^3P_2$ excited state of xenon \cite{Player1970},
as well as with several molecules, 
see, e.g., Refs.~\cite{Hudson2011,Kara2012,TheACMECollaboration2014,DeMille2013}.	
Most recently, EDM measurements in molecules with {\P}- and {\T}-odd nuclear magnetic quadrupole moments have also been proposed \cite{FlambaumMQM2014}.


For static effects, only measurements of static PNC amplitudes from conventional PNC experiments are needed to place limits on the cosmic-field parameters.
Data from such experiments already exist for some systems.
For the dynamic effects, however, a completely different style of experiment, in which one would measure small oscillations in the PNC amplitude or atomic EDM, is needed. 
The frequency and amplitude of these oscillations would enable one to extract values (or at least limits) on the relevant field parameters~\cite{Graham2011,Graham2013,Budker2014}. 
For example, if we consider an axion field, a determination of the frequency of oscillations would lead directly to a value for the mass of the particle. 
Combined with this information, the amplitude of these oscillatory effects would lead to a determination of the constants $\eta$, $\zeta$, or $b_0$.

The frequencies of the dynamic effects induced by pseudoscalar and pseudovector fields are determined (primarily) by the masses of the particles associated with these fields. 
These masses cannot be predicted in an {\sl ab initio} manner from existing theory and, as such, we treat them as independent variables in the present work. 
In the case of axions, the ``classical'' region 
($10^{-6}$ -- $10^{-4}$ eV) 
and the ``anthropic'' region 
($10^{-10}$ -- $10^{-8}$ eV) 
are regarded as two of the more likely windows in which the axion mass may lie, see, e.g.,~\cite{Kim2010}.
Axions lying in the classical or anthropic regions would lead to oscillations with frequencies on the order of GHz and MHz, respectively. 
For the case of axions, the coherence time,
$\tau_c\sim 2\pi/m_av^2$,
may be estimated from 
$\Delta \omega_a/ \omega_a\sim(\tfrac{1}{2}m_av^2/m_a)\sim v^2$, 
where a virial velocity of $v \sim 10^{-3}$ would be typical in our local Galactic neighborhood, and $\omega_a\approx m_a$~\cite{Budker2014,Graham2011}.

In the case of an axion field, with the assumption that axions saturate the CDM density of the galaxy,
the coefficients in (\ref{eq:lps}) can be recognized as $\eta=\zeta=a_0/f_a\sim 4 \times 10^{-18}$
', see, e.g., Ref.~\cite{Stadnik2014}.
For the PS fields presented in Eq.~(\ref{eq:lps}), this leads to oscillating atomic EDMs with magnitudes on the order of $10^{-38}\,e\cdot{\rm cm}$.

It is also possible to gain a further enhancement in the sensitivity of the EDM measurements, see, e.g., Refs.~\cite{Graham2011,Graham2013,Budker2014,Stadnik2014}, where oscillating EDM experiments have been recently considered. 
This can be achieved by tuning the experiment to a specific frequency in order to bring about a resonance, with 
$(E_a-E_n)^2\simeq\omega_\phi^2$, 
see Eq.~(\ref{eq:cnt}).
Similar techniques have already been shown to work using the practically degenerate $A$ and $B$ states in dysprosium~\cite{Nguyen1997}, and could potentially be implemented in systems such as barium, radium, thorium, and singly-ionized actinium, which also possess pairs of very close levels of opposite-parity~\cite{RobertsCL2014}.

\section{Conclusion}\label{sec:concl}

We have performed relativistic calculations of parity nonconservation amplitudes and atomic electric dipole moments induced by the interaction of pseudoscalar and pseudovector cosmic fields with atomic electrons for 
H, Li, Na, K, Cu, Rb, Ag, Cs, Ba, Ba$^+$, Dy, Yb, Au, Tl, Fr, and Ra$^+$.
We have shown that a static pseudoscalar cosmic field cannot give rise to observable {\P}-odd effects in atoms in the lowest order, but in contrast, a static pseudovector cosmic field can.  
Candidates for such cosmic fields include dark matter (such as axions) and dark energy, as well as a number of more exotic sources, e.g.~those described by Lorentz-invariance violating standard-model extensions~\cite{Colladay1998}.

For the case of a static pseudovector field, these calculations can be combined with existing parity nonconservation measurements to extract $1\sigma$ limits on the strength of the electron--cosmic-field coupling. 
From existing data and calculations, we find that dysprosium gives the most stringent limit for the interaction strength between the temporal component of the pseudovector field and the atomic electrons: $| b_0^e | < 7 \E{-15}$ GeV in the laboratory frame of reference. 
Also, using the existing measurement of the nuclear anapole moment of cesium and the limit on the value of the thallium nuclear anapole moment, in conjunction with their respective theoretically predicted values, we extract limits on the strength of the proton--cosmic-field couplings $b_0^p$ and  $d_{00}^p$.
By taking into account nuclear many-body effects~\cite{StadnikNMB2014}, we also extract $1\sigma$ limits on the strength of the neutron--cosmic-field couplings.  
We find that the more stringent limits of 
$\left|b_0^p \right| < 4\E{-8}$ GeV
and
$\left|d_{00}^p \right| < 5\E{-8}$ 
for protons, and
$\left|b_0^n \right| < 2 \E{-7}$ GeV
and
$\left|d_{00}^n \right| < 2 \E{-7}$ 
for neutrons come from the anapole moment results for cesium. 
These limits on the temporal components $b_0$, which are derived from {\P}-odd fermion effects, are complementary to the existing limits on the interaction of the spatial components $\v{b}$ of a static PV field with electrons, protons, and neutrons, which are derived from the {\P}-even fermion effects, see, e.g., Ref.~\cite{Kostelecky2011a,*Kostelecky2014}.

Finally, we mention that cosmic-field searches need not be restricted only to atomic systems.
Searches for cosmic-field--induced electric dipole moments can also be performed in solid-state systems. 
Static electron electric dipole moment experiments in ferroelectrics are discussed in Refs.~\cite{Eckel2012,Budker2006}, for instance, and solid-state systems have already been proposed for use in the detection of axion dark matter (see, e.g., Refs.~\cite{Budker2014,Beck2013}).
We also mention that transient electric dipole moments may also be induced by cosmic fields in the form of topological defects~\cite{StadnikDefects2014}.

\acknowledgments
The authors would like to acknowledge
Michael Hohensee, 
Iosif B.~Khriplovich, 
Derek Jackson Kimball, 
Mikhail Kozlov, 
Maxim Pospelov, 
Arkady Vainshtein, 
and 
Vladimir G.~Zelevinsky
for valuable discussions.
We are particularly grateful to V.~Alan Kosteleck\'y for pointing out that our methods in Ref.~\cite{RobertsCosmic2014} could be extended to place constraints on the $d_{00}^{\,p}$ parameter.
This work was supported in part by the Australian Research Council, by NSF Grant No.~PHY-1068875, and by the Perimeter Institute for Theoretical Physics. 
Research at the Perimeter Institute is supported by the Government of Canada through Industry Canada and by the Province of Ontario through the Ministry of Economic Development \& Innovation. 
V.~V.~Flambaum would also like to acknowledge the Humboldt foundation for support in the form of the Humboldt Award, and the MBN Research Center, where part of this work was conducted, for hospitality.
N.~Leefer was supported by a Marie Curie International Incoming Fellowship within the 7{th} European Community Framework Programme.


\bibliography{references-cosmic,other}

\end{document}